\documentclass[fleqn,usenatbib]{mnras}
\usepackage{newtxtext,newtxmath}
\usepackage[T1]{fontenc}
\DeclareRobustCommand{\VAN}[3]{#2}
\let\VANthebibliography\thebibliography
\def\thebibliography{\DeclareRobustCommand{\VAN}[3]{##3}\VANthebibliography}

\usepackage{graphicx}	
\usepackage{amsmath}	
\usepackage{tikz}
\usetikzlibrary{arrows.meta,positioning,shapes.geometric}

\title[Numerical stability of 21-cm calibration]{Impact of numerical stability in Bayesian noise wave calibration on global 21-cm experiments}

\author[Dasgupta et al.]{
Saswata Dasgupta,\textsuperscript{\href{https://orcid.org/0000-0001-6461-769X}{\includegraphics[width=2.5mm]{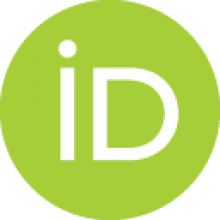}}}$^{1,2}$\thanks{E-mail: saswata.iiti@gmail.com}
Adarsh Kumar Dash,\textsuperscript{\href{https://orcid.org/0000-0002-7544-7881}{\includegraphics[width=2.5mm]{orcid.png}}}$^{1,3}$
Dominic Anstey,$^{1,3}$
Harry T. J. Bevins,$^{1,3}$
\newauthor
~Christian Kirkham,\textsuperscript{\href{https://orcid.org/0000-0001-5385-6329}{\includegraphics[width=2.5mm]{orcid.png}}}$^{1,3}$
Eloy de Lera Acedo$^{1,3}$
\\ 
$^{1}$Kavli Institute for Cosmology in Cambridge, University of Cambridge, Madingley Road, Cambridge, CB3 0HA, UK\\
$^{2}$Institute of Astronomy, University of Cambridge, Madingley Road, Cambridge, CB3 0HA, UK\\
$^{3}$Astrophysics Group, Cavendish Laboratory, University of Cambridge, J. J. Thomson Avenue, Cambridge, CB3 0US, UK\\}

\def \hi {H\textsc{i}}

\date{Accepted XXX. Received YYY; in original form ZZZ}

\pubyear{\the\year{}}

\begin{document}

\label{firstpage} 
\pagerange{\pageref{firstpage}--\pageref{lastpage}} 

\maketitle  

\begin{abstract}
Detecting the global 21-cm signal from the Cosmic Dawn and Epoch of Reionization requires calibration accuracy far below the level of astrophysical foregrounds. REACH models its receiver using the noise wave formalism, with five frequency-dependent low-noise amplifier parameters fitted jointly to multiple calibration sources. We identify a numerical instability in this Bayesian calibration pipeline: the condition number of the posterior covariance matrix reaches $\kappa(\mathbf{V}^*) \sim 10^{9}$--$10^{11}$, making solutions non-reproducible across computing environments. Singular value decomposition shows that the instability is driven by near-collinearity between the design-matrix columns associated with the excess noise source temperature, $X_\mathrm{NS}$, and the load temperature, $X_\mathrm{L}$. Using a Chebyshev basis, we develop a two-step mitigation. First, fixing $T_\mathrm{NS}$ to a scalar removes the degeneracy and reduces $\kappa(\mathbf{V}^*)$ to $\sim 60$. Second, to retain frequency dependence, we recover $T_\mathrm{NS}(\nu)$ directly from the hot-load calibration measurement. On mock data, this method preserves the stability of the reduced model while achieving comparable calibration accuracy. Masking narrow channels around cable standing-wave degeneracies further removes local artefacts in the design matrix. These steps provide a stable, reproducible, and data-driven calibration procedure. Because the $X_\mathrm{NS}$--$X_\mathrm{L}$ degeneracy is inherent to the noise wave formalism, the method is relevant to other global 21-cm experiments.

\end{abstract}

\begin{keywords}
methods: numerical -- methods: statistical -- methods: observational -- cosmology: observations -- cosmology: dark ages, reionization, first stars
\end{keywords}


\section{Introduction}

Radio emission 
from neutral hydrogen (\hi) during the early phases of the Universe, spanning the 
\textit{Cosmic Dark Ages}, \textit{Cosmic Dawn} (CD), and the \textit{Epoch of Reionization} (EoR) can provide us with a plethora of information on the cosmic history. 
This emission arises from the hyperfine transition of neutral hydrogen (\hi\,) at a rest frequency 
of $\nu \simeq 1420~\mathrm{MHz}$ (corresponding to a wavelength of $\lambda \simeq 21~\mathrm{cm}$), 
and is commonly measured relative to 
the temperature of the cosmic microwave background (CMB; \citealt{furlanetto06}). 
By observing the evolution of this differential brightness temperature, it becomes 
possible to infer the physical conditions of the intergalactic medium (IGM) and to constrain 
the astrophysical properties of the first luminous sources, including the formation of the 
earliest galaxies and the influence of dark matter during the 
Cosmic Dawn (\citealt{monslave17, monslave18, bevins22}).

Single antenna experiments aim to measure this cosmological 21-cm signal averaged over the whole sky, therefore called the global 21-cm experiments. These include experiments such as EDGES \citep{bowman18}, SARAS \citep{singh22}, REACH \citep{eloy22}, MIST \citep{monslave24}, RHINO \citep{bull24}, PRI$^\mathrm{Z}$M \citep{philip19}, LEDA \citep{price18}, GINAN \citep{mckay26} and EIGSEP \citep{bye26}.

Despite the ongoing observational effort, the detection of the cosmological 21-cm signal is extremely difficult due to its faint nature, particularly in comparison to the bright Galactic and extragalactic foregrounds which are expected to be $10^3-10^4\,\rm K$ \citep{shaver99, datta10}. Additional hurdles specific to global-signal experiments include antenna beam chromaticity \citep{spinelli22}, the effect of the ionosphere \citep{datta16,shen21}, calibration error \citep{sims20}, and the horizon and ground-screen environment \citep{pattinson24}.

The first potential detection of the global 21-cm signal was reported by the \textsc{EDGES} experiment \citep{bowman18}, which measured a flattened Gaussian absorption feature with a depth of $500^{+500}_{-200}\,\mathrm{mK}$ centred at $78 \pm 1\,\mathrm{MHz}$. The unusual depth and shape of this feature, however, require the invocation of exotic physics to explain it \citep{barkana18, feng18}. More recent observations by the \textsc{SARAS} experiment have placed stringent constraints on this signal, ruling out the \textsc{EDGES} detection at $95.3\%$ confidence \citep{singh22}.  

Several studies have raised concerns that unmodelled systematics in the \textsc{EDGES} data may be responsible for the claimed signal \citep{hills18b, sims20}. \citet{hills18b} showed that when the \textsc{EDGES} foreground model is compared with a physically motivated description, it yields non-physical parameter values, and that the fit improves significantly upon inclusion of a $12.5\,\mathrm{MHz}$ sinusoidal term without a signal profile. This interpretation is supported by \citet{sims20}, who argue that such a sinusoidal feature could arise from a calibration artefact.

The global 21-cm absorption signal is expected to be extremely faint, with an amplitude of $\lesssim 200\,\mathrm{mK}$ \citep{dhanda25b}. Therefore, it is essential to achieve highly precise instrument calibration. Small frequency-dependent systematics can easily dominate over the cosmological signal if left uncorrected. In this work, we investigate the numerical stability of the REACH Bayesian noise wave calibration pipeline \citep{roque21}. We adopt a conjugate prior framework to fit for the frequency-dependent noise wave parameters (NWPs) using mock data based on the REACH system design, and identify a critical ill-conditioning problem that renders solutions non-reproducible across computing environments. We present diagnostics based on singular value decomposition (SVD) and the condition number of the posterior covariance matrix, and propose a physically motivated mitigation strategy.

In Section~\ref{sec:methods} we describe the REACH receiver system, the noise wave parametrisation, the Bayesian conjugate prior framework, and the mock data generation procedure. In Section~\ref{sec:motivation}, we present the numerical instability problem, diagnose its origin through SVD and condition number analysis, and describe the mitigation strategies we employ. In Section~\ref{sec:results} we validate the mitigated pipeline on mock data, and in Section~\ref{sec:conclusions} we summarise our findings.

\section{Methods}
\label{sec:methods}
\subsection{Antenna Temperature and Signal Chain}

The REACH instrument is designed to measure the sky-averaged 21-cm signal from the CD-EoR by capturing the antenna temperature
over a wide band of frequencies \citep[$50-130$~MHz]{roque25}. The induced voltage at the antenna terminals is amplified
and conditioned by a receiver and recorded as power spectral density (PSD) measurements. 
The power received at frequency $\nu$ from a sky temperature distribution $T_\mathrm{sky}(\Omega, \nu, t)$, weighted by the antenna directivity $B(\Omega, \nu)$, is given by:
\begin{equation}
    T_\mathrm{ant}(\nu, t) = \int T_\mathrm{sky}(\Omega, \nu, t) B(\Omega, \nu) \mathrm{d}\Omega + T_\mathrm{rec}(\nu),
\end{equation}
where $T_\mathrm{rec}(\nu)$ is the equivalent antenna temperature of the mean additive power from noise generated internally in the LNA, as described by the noise-wave model. The underlying noise is stochastic; calibration estimates its mean contribution. Random fluctuations about that mean (the radiometric noise) cannot be subtracted; their uncertainty is instead reduced by averaging and propagated to the calibrated temperature (Section~\ref{sec:results}). The quantity of interest is the global 21-cm signal $T_{21}(\nu)$, which in principle can be isolated once the foreground contribution and the receiver's mean contribution are known:
\begin{equation}
    T_{21}(\nu) \approx T_\mathrm{ant}(\nu, t) - \int T_\mathrm{f}(\Omega, \nu, t) B(\Omega, \nu) \mathrm{d}\Omega - T_\mathrm{rec}(\nu).
\end{equation}
In practice, neither the foreground integral nor $T_\mathrm{rec}(\nu)$ is known a priori: the former must be estimated through foreground modelling, and the latter through calibration of the receiver's noise-wave parameters (Section~\ref{sec:NWP_modelling}). The recovery of $T_{21}$ therefore requires precise calibration of the instrument.

\subsection{Noise Wave Modeling} \label{sec:NWP_modelling}

The calibration of an instrument maps the measured PSDs to the temperature of a source by correcting all effects of the instrument.
To calibrate the REACH instrument, we employ the \textit{Dicke switching} technique
originally proposed by \citet{dickie46}, and the noise-wave formalism developed by \citet{meys78}. 
These have been earlier adapted for global
21-cm experiments by \citet{rogers12}. In this method, two reference sources,
a resistive load and a noise diode, are alternately measured to determine the
system gain and noise characteristics. To mitigate impedance mismatches within
the receiver chain, twelve additional reference sources are used, which are
selected to maximise coverage across the Smith chart. This enables accurate
characterisation of unknown sources such as the antenna \citep{roque21}.

The twelve reference sources employed in the REACH receiver are listed below. Eleven of these are used as calibrators in the fit; the twelfth, the $91\,\Omega$ ambient load connected through the 2-metre cable (`c2r91'), is deliberately withheld from every calibration fit and used only as a validation source, as described below.
\begin{itemize}
    \item An ambient $50\,\Omega$ `cold' load;
    \item Ambient $25\,\Omega$ and $100\,\Omega$ loads;
    \item A heated $50\,\Omega$ `hot' load maintained at $370\,\mathrm{K}$ and connected via a 4-inch cable;
    \item $27\,\Omega$, $36\,\Omega$, $69\,\Omega$, and $91\,\Omega$ ambient loads each connected through a 2-metre `short' cable;
    \item $10\,\Omega$, $250\,\Omega$, open, and short loads connected via a 10-metre `long' cable.
\end{itemize}

\begin{figure*}
    \includegraphics[width=\textwidth]{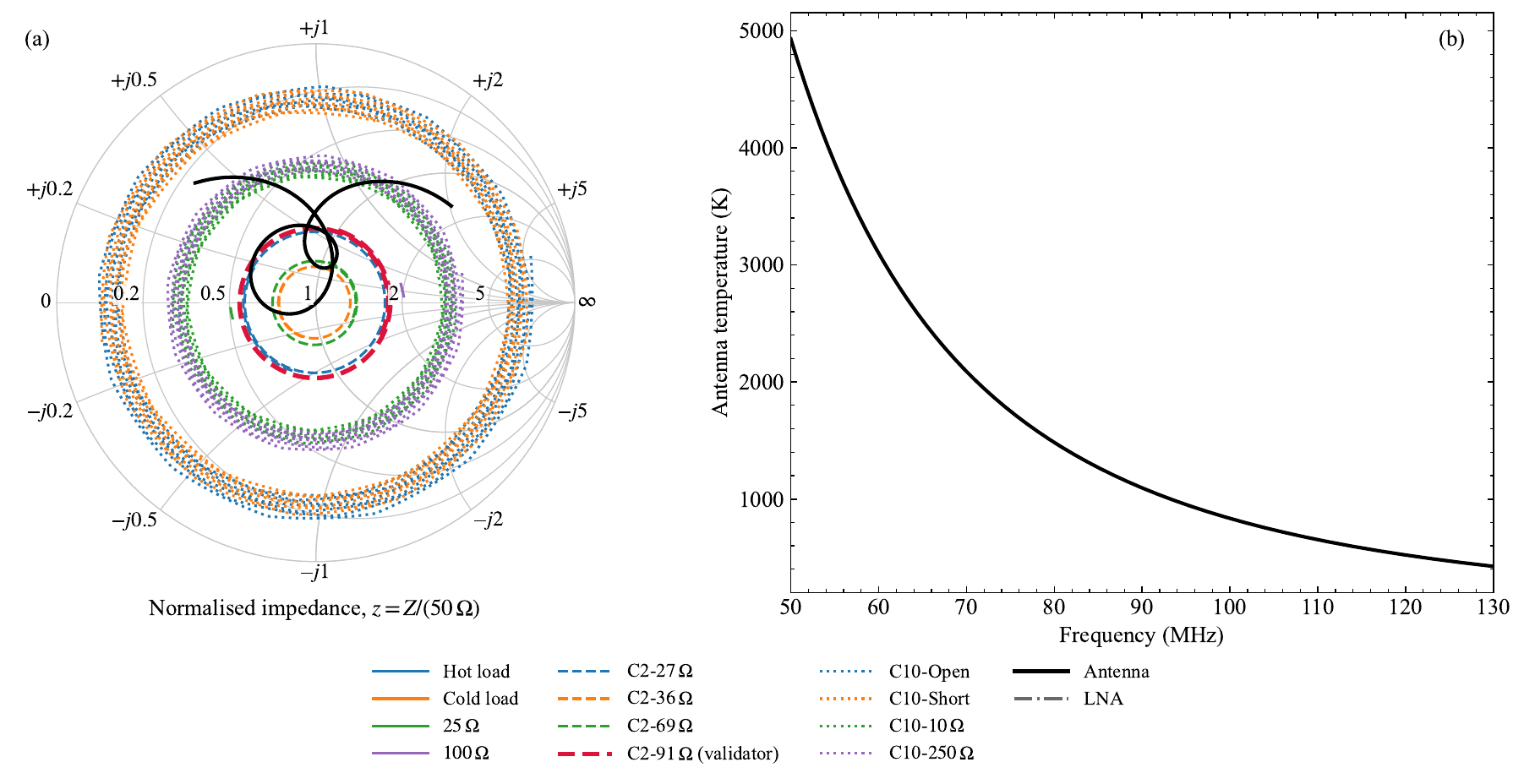}
    \caption{Left (a): Complex reflection coefficients of the twelve reference sources, the antenna, and the LNA over $50$--$130$~MHz, plotted on a Smith chart with a $50\,\Omega$ reference impedance. The chart labels give normalised resistance and reactance, $z=Z/(50\,\Omega)=r+jx$, so $r=1$ corresponds to $50\,\Omega$. Line styles distinguish uncabled loads (solid), sources on the 2-metre cable (dashed), and sources on the 10-metre cable (dotted); colours distinguish sources within each group. The validator c2r91 is the thick crimson dashed curve, the antenna is black, and the LNA is dash-dotted grey. Right (b): Antenna temperature of the mock data used in this work. Source temperatures, which are frequency-independent in this mock realisation, are listed in Table~\ref{tab:mock_temperatures} instead of being plotted.}
    \label{fig:s11_temp_mock4}
\end{figure*}

Figure~\ref{fig:s11_temp_mock4} shows the reflection coefficients of the different sources for the mock data used in this work.
`c10' refers to the elements connected through the 10-metre cable, and `c2' refers to the elements
that are connected through the 2-metre cable. The corresponding source temperatures, which are frequency-independent in this mock realisation, are listed in Table~\ref{tab:mock_temperatures}. Additionally, in our work, we use an internal $91 \Omega$ ambient load connected through a 2-metre cable to validate our calibration methods.
This choice is preferred as we require a source with a chromatic reflection coefficient, as in the antenna, to validate the constrained noise wave parameters. This chromaticity lies mainly in the phase of $\Gamma$. In the left panel of Figure~\ref{fig:s11_temp_mock4}, c2r91 traces a nearly circular path about the chart centre, showing phase variation at nearly constant reflection magnitude. Its phase changes by approximately $1.28$ cycles over $50$--$130$~MHz in the underlying S-parameter data, driven by the electrical delay of the 2-metre cable. It is this phase structure, entering equations~\eqref{eq:x_cos} and~\eqref{eq:x_sin} through $\Gamma_\mathrm{cal}$, that makes c2r91 a suitable validator for the cosine and sine noise-wave terms. As the validator is omitted from the calibration set, the choice should not deteriorate the quality of the calibration. The c2r91 source has been shown to satisfy this condition, primarily using condition number analysis in \cite{dash26}. Henceforth, we will be interchangeably calling this load `c2r91' or the `validator' in this work. The validator shall be removed in the calibration solution tests, where the remaining 11 sources will be used as calibrators. The antenna itself is a hexagonal dipole \citep{cumner22}, connected to the receiver through a 1-metre cable.

\begin{table}
    \centering
    \caption{Physical/termination temperatures of the twelve reference sources for the mock dataset used in this work (Section~\ref{sec:NWP_modelling}). Every source other than the hot and cold loads is fixed near ambient temperature; all values are frequency-independent to within the precision of the mock realisation.}
    \label{tab:mock_temperatures}
    \begin{tabular}{lc}
        \hline
        Source & Temperature (K) \\
        \hline
        Hot load & $370.00$ \\
        Cold load & $320.00$ \\
        $25\,\Omega$ & $321.62$ \\
        $100\,\Omega$ & $319.39$ \\
        C2-$27\,\Omega$ & $315.90$ \\
        C2-$36\,\Omega$ & $315.90$ \\
        C2-$69\,\Omega$ & $315.90$ \\
        C2-$91\,\Omega$ (validator) & $315.90$ \\
        C10-Open & $318.14$ \\
        C10-Short & $318.14$ \\
        C10-$10\,\Omega$ & $318.14$ \\
        C10-$250\,\Omega$ & $318.14$ \\
        \hline
    \end{tabular}
\end{table}

For each calibrator, we record the Power Spectral Density (PSD) of the calibration source, $P_\mathrm{cal}$, the reference load, 
$P_\mathrm{L}$, and the noise source, $P_\mathrm{NS}$, using a spectrometer after amplification by a receiver chain. 
We also measure the reflection coefficient of each calibration source at the receiver input, $\Gamma_\mathrm{cal}$, 
and the reflection coefficient of the receiver, $\Gamma_\mathrm{rec}$, using a vector network analyser (VNA). The standard uncalibrated antenna temperature from this process is:
\begin{equation}
    T^*_\mathrm{cal} = T_\mathrm{NS} \frac{P_\mathrm{cal} - P_\mathrm{L}}{P_\mathrm{NS} - P_\mathrm{L}} + T_\mathrm{L},
\end{equation}
where $T_\mathrm{NS}$ and $T_\mathrm{L}$ are the excess noise temperature of the noise source and the physical temperature of the cold load, respectively.

However, the presence of impedance mismatches necessitates the use of a full noise wave model. The output power can be written as \citep{rogers12}:
\begin{equation}
\begin{split}
P_\mathrm{cal} & = g_\mathrm{sys} \Bigg[ T_\mathrm{cal} (1 - |\Gamma_\mathrm{cal}|^2) \left| \frac{\sqrt{1 - |\Gamma_\mathrm{rec}|^2}}{1 - \Gamma_\mathrm{cal} \Gamma_\mathrm{rec}} \right|^2 \\
& + T_\mathrm{unc} |\Gamma_\mathrm{cal}|^2 \left| \frac{\sqrt{1 - |\Gamma_\mathrm{rec}|^2}}{1 - \Gamma_\mathrm{cal} \Gamma_\mathrm{rec}} \right|^2  + T_\mathrm{cos}\, \Re\left( \Gamma_\mathrm{cal} \frac{\sqrt{1 - |\Gamma_\mathrm{rec}|^2}}{1 - \Gamma_\mathrm{cal} \Gamma_\mathrm{rec}} \right) \\
& + T_\mathrm{sin}\, \Im\left( \Gamma_\mathrm{cal} \frac{\sqrt{1 - |\Gamma_\mathrm{rec}|^2}}{1 - \Gamma_\mathrm{cal} \Gamma_\mathrm{rec}} \right) + T_0 \Bigg],
\end{split}
\end{equation}

where $T_\mathrm{unc}$, $T_\mathrm{sin}$ and $T_\mathrm{cos}$ are the uncorrelated, sine and cosine components of the noise wave, respectively, and $g_\mathrm{sys}$ is the gain response of the receiver.
After dividing through by the system gain and rearranging, this motivates a linear model of the form
\begin{equation}
    T_\mathrm{cal} = X_\mathrm{unc}\, T_\mathrm{unc} + X_\mathrm{cos}\, T_\mathrm{cos} + X_\mathrm{sin}\, T_\mathrm{sin} + X_\mathrm{NS}\, T_\mathrm{NS} + X_\mathrm{L}\, T_\mathrm{L} + \sigma_\mathrm{cal},
    \label{eq:linear_model}
\end{equation}
with a measurement noise term $\sigma_\mathrm{cal}$ and the design-matrix elements $X_i$ are constructed from the measured
reflection coefficients and PSDs as follows:
\begin{align}
    X_\mathrm{unc} &= -\frac{|\Gamma_\mathrm{cal}|^2}{1 - |\Gamma_\mathrm{cal}|^2}, \label{eq:x_unc}\\[4pt]
    X_\mathrm{L} &= \frac{|1 - \Gamma_\mathrm{cal}\Gamma_\mathrm{rec}|^2}{1 - |\Gamma_\mathrm{cal}|^2}, \label{eq:x_l}\\[4pt]
    X_\mathrm{cos} &= -\Re\!\left( \frac{\Gamma_\mathrm{cal}}{1 - \Gamma_\mathrm{cal} \Gamma_\mathrm{rec}} \times \frac{X_\mathrm{L}}{\sqrt{1 - |\Gamma_\mathrm{rec}|^2}} \right), \label{eq:x_cos}\\[4pt]
    X_\mathrm{sin} &= -\Im\!\left( \frac{\Gamma_\mathrm{cal}}{1 - \Gamma_\mathrm{cal} \Gamma_\mathrm{rec}} \times \frac{X_\mathrm{L}}{\sqrt{1 - |\Gamma_\mathrm{rec}|^2}} \right), \label{eq:x_sin}\\[4pt]
    X_\mathrm{NS} &= \left( \frac{P_\mathrm{cal} - P_\mathrm{L}}{P_\mathrm{NS} - P_\mathrm{L}} \right) X_\mathrm{L}. \label{eq:x_ns}
\end{align}
Here $\boldsymbol{\Theta} = (T_\mathrm{unc}, T_\mathrm{cos}, T_\mathrm{sin}, T_\mathrm{NS}, T_\mathrm{L})$ is the vector of five frequency-dependent noise wave parameters to be fitted.
Details on the derivation of this linear model can be found in \cite{roque21}.

\subsection{Bayesian Conjugate Priors Framework}
\label{sec:conjugate_priors}

In this work, we employ a Bayesian approach using conjugate priors as described in \cite{roque21} to fit for the noise wave parameters $\boldsymbol{\Theta}$. To allow for smooth frequency dependence, each noise wave parameter is promoted to a vector of polynomial coefficients $T_i = (T_i^{[0]}, T_i^{[1]}, \ldots, T_i^{[n_i]})$, where $n_i$ is the polynomial order for the $i$-th parameter. The design matrix $\mathbf{X}$ is likewise expanded so that the linear model in equation~\eqref{eq:linear_model} becomes $\mathbf{T}_\mathrm{cal} = \mathbf{X}\boldsymbol{\Theta} + \boldsymbol{\sigma}$, where $\mathbf{T}_\mathrm{cal}$ is a vector over frequency and $\boldsymbol{\sigma}$ represents the measurement noise.

Assuming a Gaussian likelihood with unknown variance $\sigma^2$, the probability of the data given the model parameters is
\begin{equation}
\begin{split}
    p(\mathbf{T}_\mathrm{cal} \mid \boldsymbol{\Theta}, \sigma^2) &= \left(\frac{1}{2\pi\sigma^2}\right)^{N/2} \\
    &\quad\times\exp\!\left(-\frac{1}{2\sigma^2}(\mathbf{T}_\mathrm{cal} - \mathbf{X}\boldsymbol{\Theta})^\top (\mathbf{T}_\mathrm{cal} - \mathbf{X}\boldsymbol{\Theta})\right),
\end{split}
    \label{eq:likelihood}
\end{equation}
where $N$ is the number of measurements.

We adopt a normal-inverse-gamma conjugate prior on the parameters and noise variance, following \cite{banerjee08} and \cite{roque21}:
\begin{equation}
\begin{split}
    p(\boldsymbol{\Theta}, \sigma^2) &\propto \left(\frac{1}{\sigma^2}\right)^{a_0 + 1 + d/2} \\
    &\quad\times\exp\!\left(-\frac{1}{\sigma^2}\left\{b_0 + \frac{1}{2}(\boldsymbol{\Theta} - \boldsymbol{\mu}_0)^\top \mathbf{V}_0^{-1} (\boldsymbol{\Theta} - \boldsymbol{\mu}_0)\right\}\right),
\end{split}
    \label{eq:prior}
\end{equation}
where $d$ is the dimensionality of $\boldsymbol{\Theta}$, $\boldsymbol{\mu}_0$ is the prior mean, $\mathbf{V}_0$ is the prior covariance matrix, and $a_0 > 0$, $b_0 > 0$ are the shape and scale hyperparameters of the inverse-gamma distribution on $\sigma^2$.

Since the prior is conjugate to the Gaussian likelihood, the posterior has the same functional form with updated hyperparameters \citep{banerjee08, roque21}:
\begin{align}
    \mathbf{V}^* &= \left(\mathbf{V}_0^{-1} + \mathbf{X}^\top \mathbf{X}\right)^{-1}, \label{eq:Vstar}\\
    \boldsymbol{\mu}^* &= \mathbf{V}^*\left(\mathbf{V}_0^{-1}\boldsymbol{\mu}_0 + \mathbf{X}^\top \mathbf{T}_\mathrm{cal}\right), \label{eq:mustar}\\
    a^* &= a_0 + \frac{N}{2}, \label{eq:astar}\\
    b^* &= b_0 + \frac{1}{2}\left(\boldsymbol{\mu}_0^\top \mathbf{V}_0^{-1}\boldsymbol{\mu}_0 + \mathbf{T}_\mathrm{cal}^\top \mathbf{T}_\mathrm{cal} - \boldsymbol{\mu}^{*\top} \mathbf{V}^{*-1}\boldsymbol{\mu}^*\right). \label{eq:bstar}
\end{align}
Here, $\boldsymbol{\mu}^*$ is the posterior mean of the noise wave polynomial coefficients and $a^*$, $b^*$ are the updated inverse-gamma hyperparameters. The posterior covariance $\mathbf{V}^*$ (equation~\eqref{eq:Vstar}) encodes how well each polynomial coefficient is constrained by the data: its diagonal entries give the marginal variances, and its off-diagonal entries capture correlations between coefficients. Overall, $\mathbf{V}^*$ combines information from both the prior ($\mathbf{V}_0^{-1}$) and the data ($\mathbf{X}^\top\mathbf{X}$), so its condition number $\kappa(\mathbf{V}^*)$ directly measures the numerical stability of the calibration solution. A large $\kappa(\mathbf{V}^*)$ implies that small perturbations in the data or numerical rounding can produce large changes in the inferred parameters, which is the central problem investigated in this paper. The marginal likelihood (evidence) is obtained by integrating over $\boldsymbol{\Theta}$ and $\sigma^2$:
\begin{multline}
    p(\mathbf{T}_\mathrm{cal}) = \int p(\mathbf{T}_\mathrm{cal} \mid \boldsymbol{\Theta}, \sigma^2)\, p(\boldsymbol{\Theta}, \sigma^2)\, \mathrm{d}\boldsymbol{\Theta}\, \mathrm{d}\sigma^2 \\
    = (2\pi)^{-N/2}\frac{b_0^{a_0}\,\Gamma(a^*)\sqrt{|\mathbf{V}^*|}}{b^{*a^*}\,\Gamma(a_0)\sqrt{|\mathbf{V}_0|}}.
    \label{eq:evidence}
\end{multline}
It is important to note that the $\Gamma$ symbol used in equation~\eqref{eq:evidence} represents the gamma function, not the reflection coefficient.
The evidence is used for model comparison, in particular for determining the optimal polynomial orders for each noise wave parameter via Bayesian model selection. We also use the Bayesian evidence as a metric to quantify the robustness of the calibration pipelines as described in sections hereafter.

\subsection{Mock Data}

We use a realistic mock dataset to test our calibration pipeline. The mock dataset used here is based on detailed electromagnetic and system-level simulations of the REACH receiver. These mock datasets are produced using lab S-parameter measurements of the sources, signal paths and LNA and simulated noise parameters \citep{haus60}. In the mock data we use to demonstrate our case, the antenna is simulated as a power law spectrum, as shown in Figure~\ref{fig:s11_temp_mock4}. In this dataset, we do not model the antenna as being connected to a cable. 
The hot and cold loads are fixed at $370$~K and $320$~K, respectively.
The remaining calibrator source and cable temperatures in this mock
realisation are fixed near ambient temperature, with values between
$315.9$ and $321.6$~K.

The power-law spectrum in this data is generated using the EDGES log-polynomial fit to the Galactic foreground power \citep{bowman18}. This is modelled as

\begin{multline}
    T_\mathrm{ant} = 1801.4 \left( \frac{\nu}{\nu_{0}} \right)^{-2.5} 
                 + 116.7 \left( \frac{\nu}{\nu_{0}} \right)^{-1.5} \\
                 - 289.8 \left( \frac{\nu}{\nu_{0}} \right)^{-0.5}
                 + 144.8 \left( \frac{\nu}{\nu_{0}} \right)^{0.5}
                 - 22.9 \left( \frac{\nu}{\nu_{0}} \right)^{1.5},
\label{eq:Tant_edges}
\end{multline}

where $\nu_{0} = 75~\mathrm{MHz}$, generated over the frequency range 
$50 \leq \nu \leq 130~\mathrm{MHz}$. The frequency range of 50--130 is chosen due to the bandpass of the actual receiver, which shall be used for all tests in this work, unless stated otherwise.
We do not consider antenna chromaticity effects, assuming an achromatic antenna gain instead.

\subsection{Cable Corrections and S-Parameter Models}

For sources including cables, we apply corrections for transmission losses and thermal gradients to the temperature of the termination, $T_\mathrm{term}$ and cable, $T_\mathrm{c}$ and get the effective source temperature at the reference plane \citep{monslave17,roque25}:
\begin{equation}
    T_\mathrm{s}^\mathrm{eff} = G_\mathrm{c} T_\mathrm{term} + (1 - G_\mathrm{c}) T_\mathrm{c},
\end{equation}
where $G_\mathrm{c}$ is the available gain, derived from the cable S-parameters and the reflection coefficient $\Gamma_\mathrm{term}$ of the termination connected to the cable:
\begin{equation}
    G_c = \frac{|S_{21}|^2 (1 - |\Gamma_\mathrm{term}|^2)}{|1 - S_{11} \Gamma_\mathrm{term}|^2(1 - |\Gamma_\mathrm{s}|^2)}.
\end{equation}

\section{Numerical Stability Analysis}
\label{sec:motivation}

\subsection{Observed reproducibility failure}
\label{sec:instability}

A fundamental requirement of any calibration pipeline is reproducibility: 
identical data and code should yield the same result regardless of the computing 
environment. We encountered a failure of this requirement when running the 
standard REACH calibration pipeline on the same mock dataset under two 
different software environments. The calibration code and input data were identical; 
the only difference was the version of the NumPy library used: NumPy~1.x, 
which uses the \texttt{OpenBLAS} linear algebra backend, and NumPy~2.x, 
that defaults to Apple \texttt{Accelerate}\footnote{\url{https://numpy.org/doc/2.0/release/1.26.0-notes.html}} \citep{harris20}\footnote{This behaviour is not specific to Apple hardware or to any
single library. The IEEE 754 standard fixes the result of each
individual floating-point operation \citep{ieee754_2019}, but different
linear algebra backends add up long sums in different
orders, so their results differ at the level of the machine precision,
$\sim10^{-16}$. When the system being solved is ill-conditioned, these
tiny differences are amplified by factors of up to $\kappa$
\citep{goldberg91,higham02}. Similar disagreements would therefore
appear on Linux between OpenBLAS \citep{wang13} and Intel MKL builds of
NumPy, between different CPUs running the same
library, or even on one machine when the number of threads changes
\citep{intel_cnr}. The NumPy~1.x/2.x comparison shown here is simply
one convenient example of a general effect.}.

\begin{figure}
    \centering
    \includegraphics[width=0.48\textwidth]{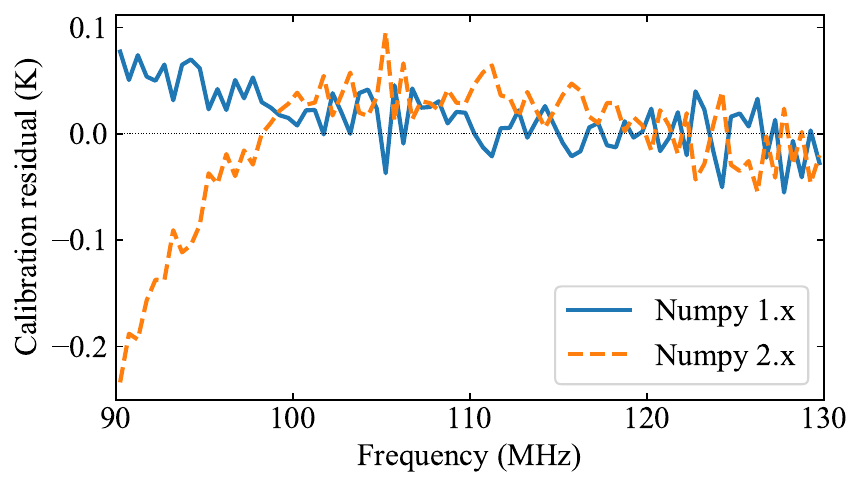}
    \caption{Calibration residuals (fitted $-$ true temperature) for source c2r91 
    under the fiducial pipeline that uses monomial basis, binned to 0.5\,MHz. The two NumPy 
    backends produce visibly different residual structure, with differences 
    reaching $\sim0.3$\,K, demonstrating that the calibration solution is not 
    reproducible across software environments.}
    \label{fig:residual_overlay}
\end{figure}

Despite running the same pipeline on the same data, the two environments 
produced quantitatively different calibration solutions. 
Figure~\ref{fig:residual_overlay} shows the calibration residuals for 
the validator source c2r91 computed under both NumPy environments: 
the two backends produce visibly different residual structure within 
the sub-band of $90-130$~MHz, with differences reaching $\sim0.3$\,K. 
In precision cosmology experiments such as global 21-cm measurements, 
where the target signal is orders of magnitude fainter than astrophysical 
foregrounds, such millikelvin-level environment-dependent errors may 
propagate directly into biased inference of the Cosmic Dawn and Epoch of 
Reionization 21-cm signal.

\begin{figure}
    \centering
    \includegraphics[width=0.48\textwidth]{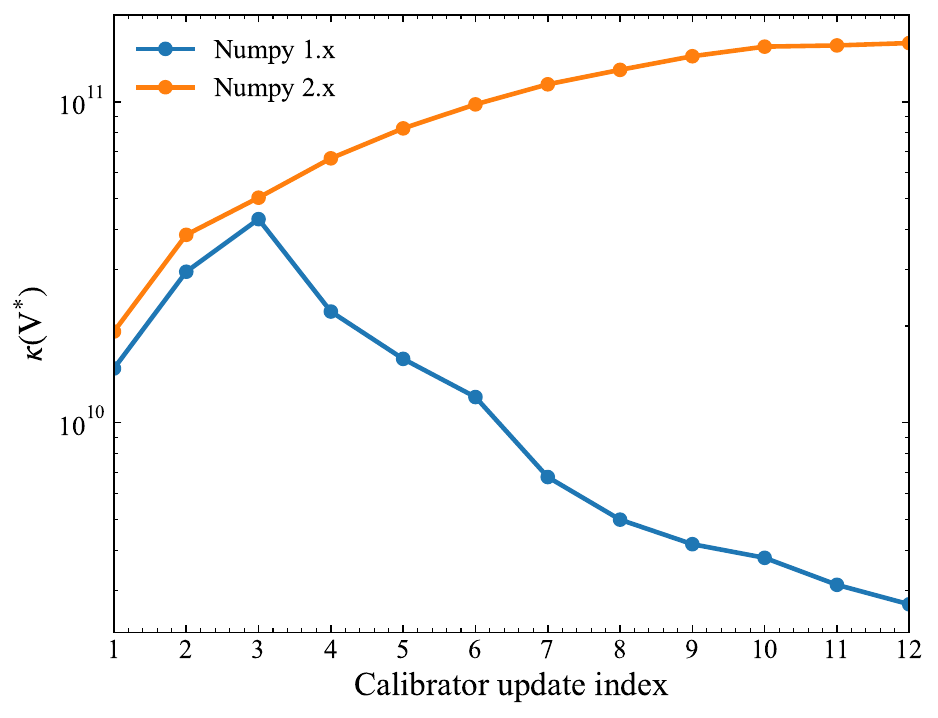}
    \caption{Condition number of the noise wave parameter posterior covariance 
    matrix, $\kappa (\mathbf{V}^*)$, as calibrators are updated sequentially, 
    computed under NumPy~1.x (\texttt{OpenBLAS}, blue) and NumPy~2.x 
    (\texttt{Accelerate}, orange) using identical mock data and calibration code. 
    The two environments yield $\kappa$ values that differ by more than an order 
    of magnitude after the final update, tracing the environment-dependent 
    residuals in Fig.~\ref{fig:residual_overlay} to an ill-conditioned posterior 
    covariance matrix.}
    \label{fig:vn_cn_pyversion}
\end{figure}

To trace the source of this discrepancy, we investigated each constituent 
of the posterior distribution (i.e.\ $a^*$, $b^*$, $\mu^*$, $\mathbf{V}^*$ as 
per \citealt{roque21} and described in equations \ref{eq:Vstar}--\ref{eq:bstar}) 
and identified the posterior covariance matrix $\mathbf{V}^*$ as the component 
responsible for the version-dependent behaviour. In a previous work, \cite{dash26} 
found that the condition number $\kappa$ is a useful diagnostic for 
quantifying the stability of the calibration solution: a high $\kappa$ 
indicates sensitivity to small perturbations in the input, and we therefore 
use this metric throughout the remainder of our analysis. Figure~\ref{fig:vn_cn_pyversion} 
shows $\kappa(\mathbf{V}^*)$ as the posteriors are updated sequentially for each calibrator under 
both NumPy environments. The two curves diverge as more calibrators are added, 
and at the end of the final update differ by more than an order of magnitude 
($\sim10^{11}$ vs $\sim10^{9}$). In most scientific computing applications a 
change in LAPACK backend would be inconsequential, since the resulting 
floating-point differences are at the level of machine precision ($\sim10^{-16}$) 
and are negligible for well-conditioned problems. When $\kappa(\mathbf{V}^*) \sim 10^{9}$--$10^{11}$, 
however, these differences are amplified by the same factor, producing calibration 
errors at the $\sim10^{-5}$ relative level. In what follows, we diagnose the two 
principal sources of the ill-conditioning and present a physically motivated 
mitigation strategy.

\subsection{Diagnosis of the instability}
\label{sec:diagnostics}

Having established that the calibration solution is environment-dependent, we now examine the structure of the posterior covariance matrix $\mathbf{V}^*$ in detail to identify the origin of the ill-conditioning. For this analysis, we fix the polynomial orders for the NWPs to values which are: 9th order for $T_\mathrm{unc}$, $T_\mathrm{cos}$, and $T_\mathrm{sin}$, and 2nd order for $T_\mathrm{NS}$ and $T_\mathrm{L}$ (whose frequency structure is well understood). We fix the orders to their highest physically reasonable value to perform this test. Otherwise, the optimal order is determined by a gradient ascent algorithm on the Bayesian evidence, following earlier works \citep{roque21,trotta08}. All subsequent diagnostics use NumPy~2.x unless otherwise stated.

\begin{figure*}
\centering
    \includegraphics[width=\textwidth]{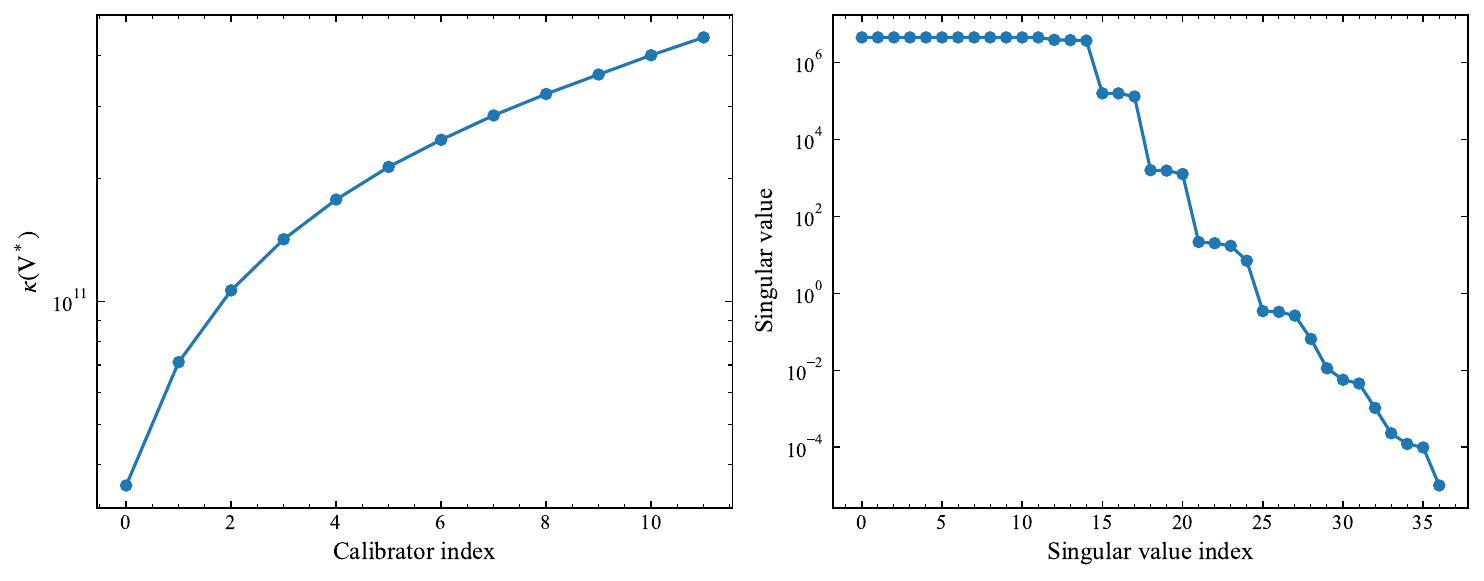}
    \caption{Left: condition number of $\mathbf{V}^*$ at each calibrator update step. Right: SVD spectrum of the final $\mathbf{V}^*$ for the same case.
    The steep drop in SVD across several decades of singular values confirms that $\mathbf{V}^*$ is nearly rank-deficient,
    indicating that the posterior is essentially constrained poorly.}
    \label{fig:pipeline_cn_svd}
\end{figure*}

Figure~\ref{fig:pipeline_cn_svd} compares the $\kappa$ and Singular Value Decompostition (SVD) of the final posterior $\mathbf{V}^*$.
The $\kappa$ (shown in the left panel) increases monotonically from $\sim 10^{10}$
after the first calibrator update and reaches to $\sim 3\times10^{11}$ after all 12 calibrator updates. The $\mathbf{V}^*$ directly depends on $\mathbf{X}^T \mathbf{X}$, where
$\mathbf{X}$ is the design matrix consisting of $X_i$ parameters as defined in Section~\ref{sec:NWP_modelling}.
The growth with each update indicates the contribution from each new calibrator to $\mathbf{X}^T \mathbf{X}$, which progressively pushes the posterior matrix closer to singularity. Additionally, a $\kappa$ of $\sim 10^{11}$ means that in double precision ($\sim 16$ significant digits), only $\sim 5$ digits of the inverse are reliable which leads to the version dependence of the solution as described previously.

A similar inference could be drawn from the SVD spectrum of the final $\mathbf{V}^*$, as shown in the right panel of Figure~\ref{fig:pipeline_cn_svd}.
We see a plateau of large singular values and then a sharp drop to a singular value of $\sim 10^{-5}$. This ``cliff" like structure
confirms that the $\mathbf{V}^*$ matrix is effectively rank-deficient. The singular values that fall below this cliff correspond to the highest-order polynomial coefficients (orders $7$--$9$ of the 9th-degree NWPs). These high-degree terms represent fine-scale frequency structure that the calibrator measurements do not have the leverage to constrain, so their associated directions in parameter space are effectively unconstrained by the calibrator data.
These near-zero singular values are the ones that get amplified during the inversion process. This leads to the $\sim 10^{11}$ condition number values, making the Cholesky decomposition sensitive to floating-point rounding.

\begin{figure*}
    \includegraphics[width=\textwidth]{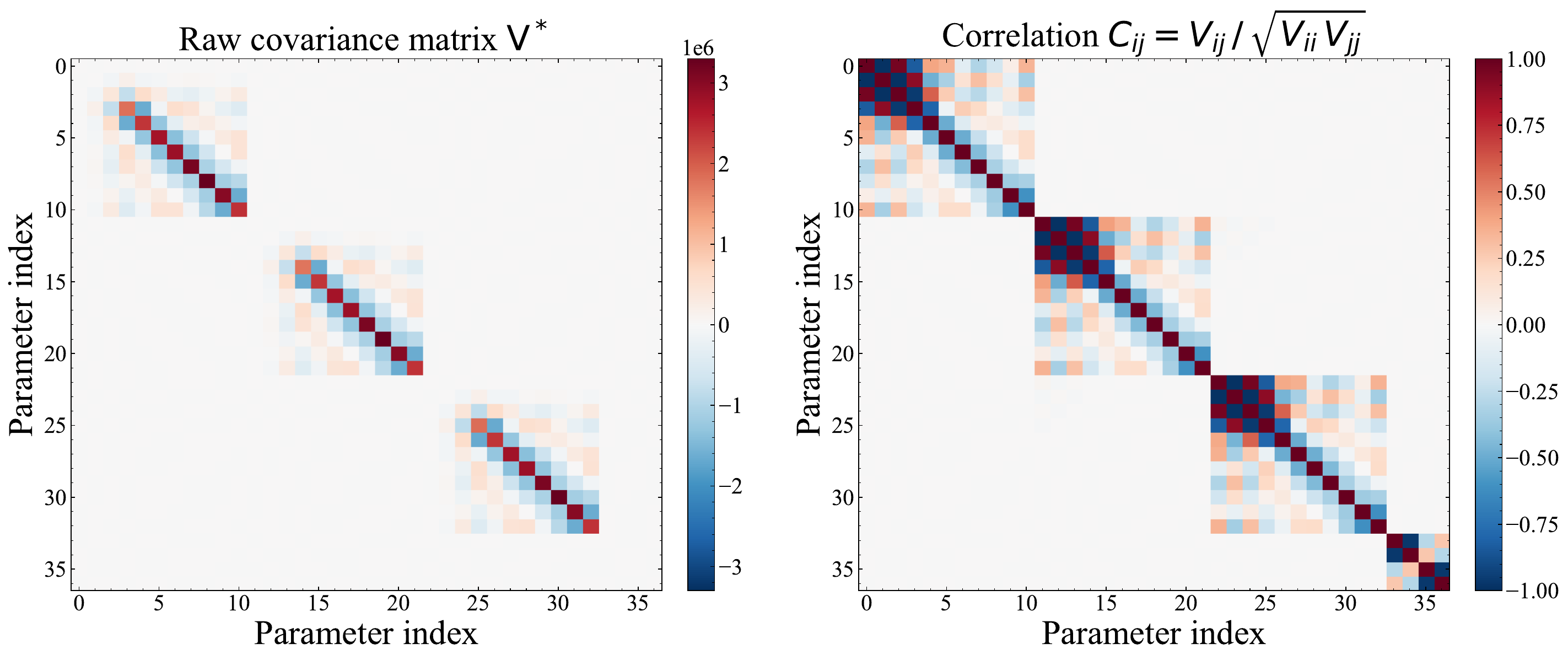}
    \caption{Left: posterior covariance matrix $\mathbf{V}^*$, where each axis indexes the polynomial coefficients of the five NWPs arranged sequentially (e.g. parameter indices $0$ to $n_1$ for $T_\mathrm{unc}$, $n_1{+}1$ to $n_1{+}n_2$ for $T_\mathrm{cos}$, and so on for $T_\mathrm{sin}$, $T_\mathrm{NS}$ and $T_\mathrm{L}$). The large dynamic range along the diagonal obscures the correlation structure.
    Right: corresponding correlation matrix $C_{ij} = \mathbf{V}^*_{ij}/\sqrt{\mathbf{V}^*_{ii}\mathbf{V}^*_{jj}}$. The strong correlations between polynomial coefficients of the same noise wave parameter (visible as bright blocks along the diagonal) show that these coefficients are not independently constrained by the data, contributing to the near-singularity of $\mathbf{V}^*$.}
    \label{fig:corr_heatmap}
\end{figure*}

To further identify and confirm the source of the ill-conditioning, we analyse 
the covariance matrix of $\mathbf{V}^*$ as shown in the left panel of 
Figure~\ref{fig:corr_heatmap}.
We observe that the colour scale of the covariance matrix (left panel) 
is dominated by large absolute values along the diagonals, reflecting the 
wide range of marginal variances across parameters. This dynamic range makes 
it difficult to visually identify the off-diagonal correlation structure. 
To isolate the correlations for visualisation purposes, we normalise the 
covariance matrix to obtain the correlation matrix (right panel), defined as 
$C_{ij} = \mathbf{V}^*_{ij}/{\sqrt{\mathbf{V}^*_{ii} \mathbf{V}^*_{jj}}}$. 
We stress that this normalisation does not alter the condition number or 
improve the conditioning; it serves only to reveal the pattern of parameter 
correlations on a common scale.

The correlation matrix reveals that the polynomial coefficients within 
the expansion of a single noise wave parameter are strongly correlated with each other. 
For example, the coefficients $\{a_0, a_1, a_2, \ldots\}$ of $T_\mathrm{unc}(\nu) = \sum_j a_j \nu^j$ 
are mutually correlated, as are those of $T_\mathrm{cos}$, $T_\mathrm{sin}$, and so on. 
These intra-parameter correlations appear as prominent blocks along the diagonal, 
with each block spanning the polynomial coefficient indices of one noise wave parameter. 
Physically, this means that the monomial basis functions $\{1, \nu, \nu^2, \ldots\}$ 
used to model the frequency dependence of each NWP become nearly collinear over the REACH band, 
so the data cannot independently constrain neighbouring polynomial coefficients.

Since $C_{ij}$ is far from an identity matrix, the ill-conditioning is structural. 
If it were purely a scaling problem (i.e.\ parameters with very different magnitudes), 
normalisation to the correlation matrix would yield something close to the identity. 
The strong off-diagonal entries that persist after normalisation confirm that the problem 
lies in the near-collinearity of the polynomial basis within each NWP.

\begin{figure*}
    \includegraphics[width=\textwidth]{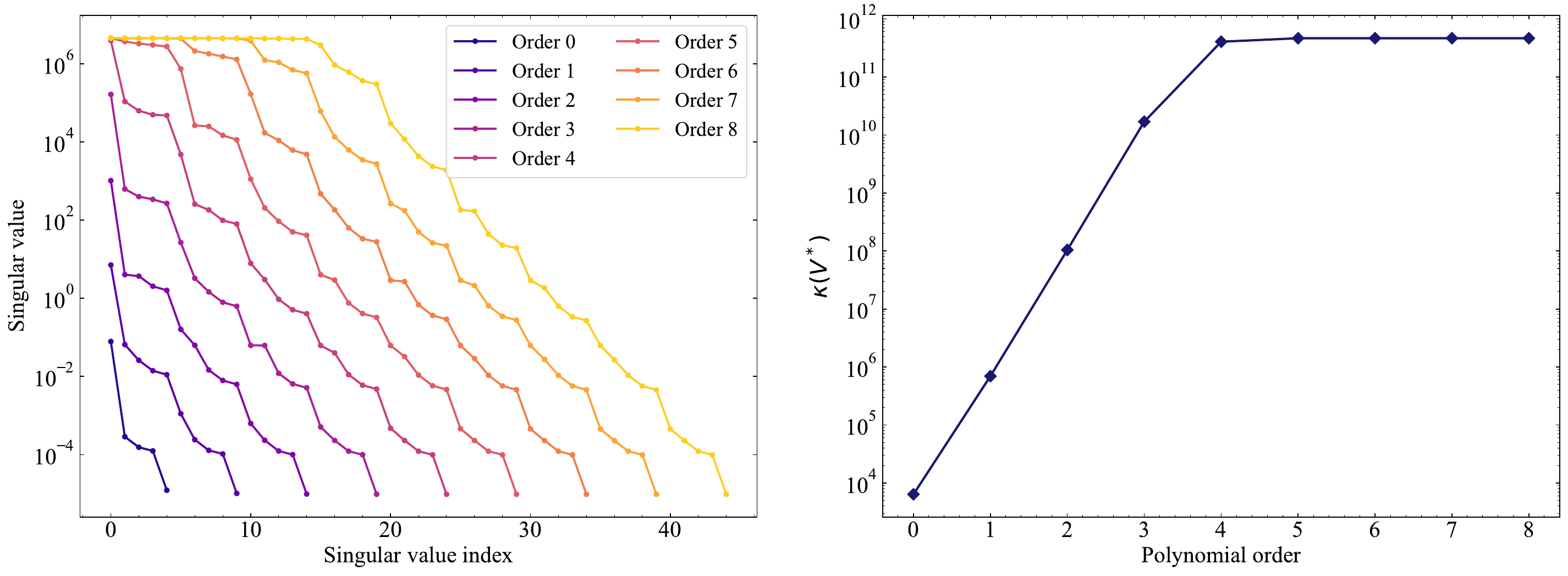}
    \caption{Left: SVD spectra for different polynomial orders. Right: Condition number ($\kappa$) as a function of polynomial order.
    The small dynamic range of singular values at low polynomial orders indicates that these coefficients are reasonably well constrained 
    by the data, while the steep drop at higher orders indicates that these coefficients are poorly constrained, contributing to the near-singularity of $\mathbf{V}^*$.}
    \label{fig:svd_orders}
\end{figure*}

Finally, we look at the SVD for increasing orders of polynomial coefficients for every NWP. To do this test, the polynomial orders were varied from 0 to 8, as it was observed that $\kappa (\mathbf{V}^*)$ saturates after order 8, as evident in Figure~\ref{fig:svd_orders}.
We find that the singular values corresponding to the higher-order polynomial coefficients (orders $4-8$) are relatively large and these contribute to the plateau in the SVD spectrum,
indicating that these coefficients are not well constrained by the data. In contrast, we observe that the singular values corresponding to the lower-order
polynomial coefficients (orders $0-3$) are quite small. The corresponding $\kappa$ values for each of these orders of polynomial coefficients are shown in the right panel of Figure~\ref{fig:svd_orders}.
Since the $\kappa$ values are defined as the ratio of the largest to the smallest singular value, we see that it monotonically increases with the order of polynomial coefficients as the dynamic
range of the singular values 
increases. This test confirms that these coefficients are poorly constrained by the data and contribute to the ill-conditioning of $\mathbf{V}^*$.
This is consistent with the fact that the data does not have enough information to constrain these high-order coefficients, leading to near-zero singular values and thus a high condition number.

\subsection{Degeneracy within the design matrix} \label{subsec:Degeneracy_in_X}

The analysis above identifies the monomial basis collinearity within each NWP as one source of ill-conditioning. We now turn to a second, independent source: a degeneracy between the design-matrix columns $\mathbf{X}_\mathrm{NS}$ and $\mathbf{X}_\mathrm{L}$.
The prior covariance $\mathbf{V_0}$ regularises the problem but is not strong enough to mitigate the ill-conditioning, confirming that it is the structure of $\mathbf{X}$ that is the root cause of the instability. Further, to reiterate, this is because the problem is extremely overdetermined, i.e, we are constraining $\sim 40$ parameters with $\sim 10^3$ of data points so $\rm X^TX$ becomes large.

We, therefore, look at the structure of each of the $X_i$ parameters and identify that
$\mathbf{X}_\mathrm{NS}$, which is the column of the design matrix corresponding to the excess noise source temperature
and the $\mathbf{X}_\mathrm{L}$, which corresponds to the load temperature, are degenerate with each other, and the data cannot tell them apart.
This is supported by the definition of these terms.
Recall from equations~\eqref{eq:x_l} and \eqref{eq:x_ns} that $X_\mathrm{NS} = X_\mathrm{L}\,(P_\mathrm{cal} - P_\mathrm{L})/(P_\mathrm{NS} - P_\mathrm{L})$.
Since the spectrometer measured PSDs for all sources are of similar order (i.e. $\sim 10^{16}$ spectrometer units), the power ratio $(P_\mathrm{cal} - P_\mathrm{L})/(P_\mathrm{NS} - P_\mathrm{L})$ is usually close to unity, causing the said degeneracy.

To test whether this is the case, we perform a controlled numerical experiment in which we replace the design-matrix columns manually.
We set $X_\mathrm{L}= 1-X_\mathrm{NS}$ and vary $X_\mathrm{NS}$ from 0 to 1, with
the rest of the $\mathbf{X}$ columns fixed to 0. We note that in the real system $X_\mathrm{NS}$ is not bounded to $[0,1]$, since the $X_\mathrm{NS}- X_\mathrm{L}$ relation depends on the power ratio; here we use this range as a normalised parametrisation for the synthetic test. As $X_\mathrm{NS}$ increases from 0 to 1, the calibrator output shifts from being dominated by the load temperature $T_\mathrm{L}$ (at $X_\mathrm{NS}=0$, the calibrator behaves as a pure resistive load) towards being dominated by the noise source temperature $T_\mathrm{NS}$ (at $X_\mathrm{NS}=1$, the output power is entirely set by the noise diode).
We then recompute $\kappa(\mathbf{V}^*)$ at increasing $X_\mathrm{NS}$.

\begin{figure}
    \centering
    \includegraphics[width=0.48\textwidth]{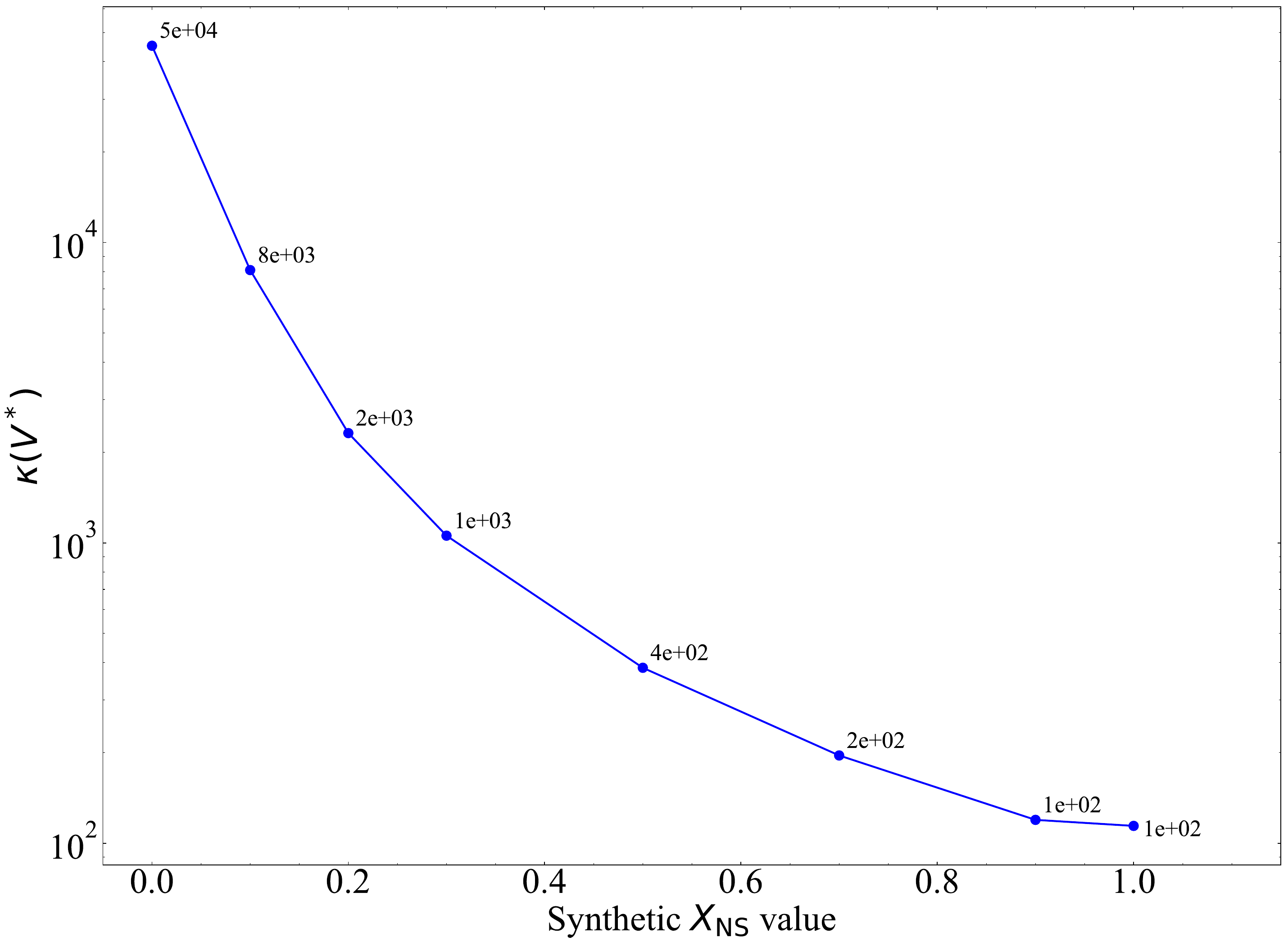}
    \caption{Condition number of $\mathbf{V}^*$ as a function of synthetic value of
    $\mathbf{X}_\mathrm{NS}$, which is the column of the design matrix corresponding to
    the noise source temperature. Given the entries $X_\mathrm{L}= 1-X_\mathrm{NS}$, the monotonic decrease in the condition number confirms that the
    degeneracy between $X_\mathrm{NS}$ and $X_\mathrm{L}$ is a major contributor
    to the ill-conditioning of $\mathbf{V}^*$.}
    \label{fig:xns_boost_cn}
\end{figure}

Figure~\ref{fig:xns_boost_cn}, confirms that the degeneracy between $X_\mathrm{NS}$ and $X_\mathrm{L}$ is the key driver of the high condition number seen in the previous pipelines.
A synthetic value of $X_\mathrm{NS} \gtrsim 0.5$
reduces $\kappa(\mathbf{V}^*)$ by several orders of magnitude relative to the
baseline. This shows that the calibration pipeline that uses all the 5 NWPs is under-constrained by the standard REACH calibrator suite.

\subsection{Mitigating the instability}
\label{sec:mitigation}

In Sections~\ref{sec:diagnostics} and \ref{subsec:Degeneracy_in_X}, 
we identified two principal sources of the ill-conditioning: 
(i) the near-collinearity of the monomial polynomial basis within each NWP, 
and (ii) the degeneracy between the $\mathbf{X}_\mathrm{NS}$ and 
$\mathbf{X}_\mathrm{L}$ columns of the design matrix. 
We now address each in turn, with the goal of reducing 
$\kappa(\mathbf{V}^*)$ to a level at which floating-point 
differences between software environments no longer affect the solution. 
A third, independent source of instability that may arise from cable $S_{11}$ 
crossovers in real REACH data is discussed separately in Section~\ref{sec:cable_masking}.

\subsubsection{The Chebyshev pipeline} \label{sec:Chebyshev_pipeline}

As a first step, we change the Bayesian update strategy from sequential to a single stacked update, where all calibrators are folded into the likelihood simultaneously.
In the sequential approach, each calibrator is used to update the posterior in turn, so the floating-point errors introduced at one step are inherited by the next as the starting prior. Over 12 calibrator updates, these errors accumulate, progressively worsening the conditioning of $\mathbf{V}^*$. Replacing this with a single stacked update breaks the error-propagation chain and all calibrators contribute to one matrix inversion, so there is no accumulated drift.
These changes reduce the $\kappa (\mathbf{V}^*)$ by a factor of $\sim 10^3$, but it is still of order $\sim 10^8$, which is too high for a stable solution.

For further improvement, we change the polynomial basis for the construction of $\mathbf{V}^*$
from monomials to Chebyshev polynomials of the first kind, $T_k(x)$. The monomial basis $\{1, \nu, \nu^2, \ldots\}$ produces columns in the design matrix $\mathbf{X}$ that become increasingly collinear at high degree, because powers of a variable over a finite interval are nearly linearly dependent. Chebyshev polynomials, by contrast, satisfy the orthogonality relation
\begin{equation}
    \int_{-1}^{1} T_m(x) T_n(x) (1 - x^2)^{-1/2}\, \mathrm{d}x = 0 \quad \mathrm{for}\ m \neq n.
\end{equation}
After mapping the REACH frequency band to the interval $[-1,1]$, the Chebyshev basis columns of $\mathbf{X}$ are therefore much less correlated, which directly reduces the condition number of $\mathbf{X}^\top\mathbf{X}$ and hence of $\mathbf{V}^*$. We collectively refer to these changes as the `Chebyshev pipeline'
and refer to the previous pipeline, without these improvements, as the `Monomial pipeline'.

\begin{figure}
    \centering
    \includegraphics[width=\columnwidth]{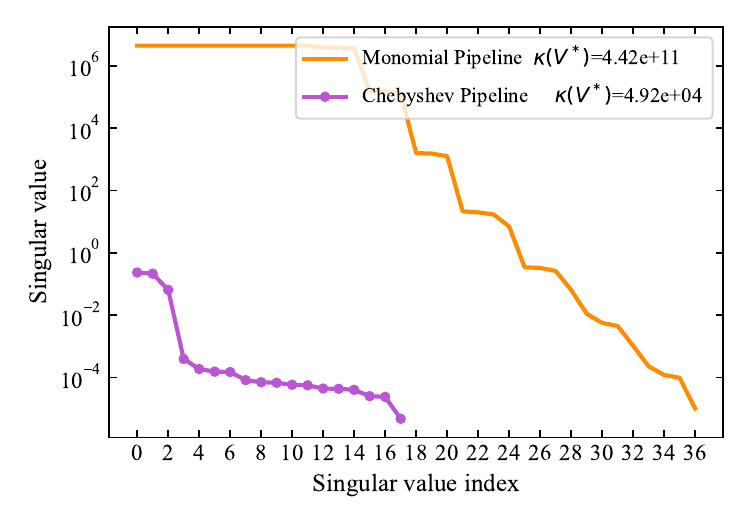}
    \caption{Comparison between the SVD spectra of $\mathbf{V}^*$ for the Monomial pipeline (orange solid) and the Chebyshev pipeline (purple dotted).
    The Chebyshev pipeline incorporates a single joint update strategy, changed priors, and a Chebyshev polynomial basis. 
    These improvements flatten the SVD spectrum and reduce $\kappa (\mathbf{V}^*) \sim 10^4$.}
    \label{fig:cheby_svd}
\end{figure}

We show the comparison between the SVD spectra of $\mathbf{V}^*$ for the Monomial pipeline in solid orange
and the Chebyshev pipeline in dotted purple in Figure~\ref{fig:cheby_svd}. After employing the
aforementioned improvements, we see that the SVD spectrum of the Chebyshev pipeline has a much
flatter profile when compared to the Monomial pipeline.
The value of $\kappa(\mathbf{V}^*)$ reduces to $\sim 10^4$.

\begin{figure*}
    \includegraphics[width=\textwidth]{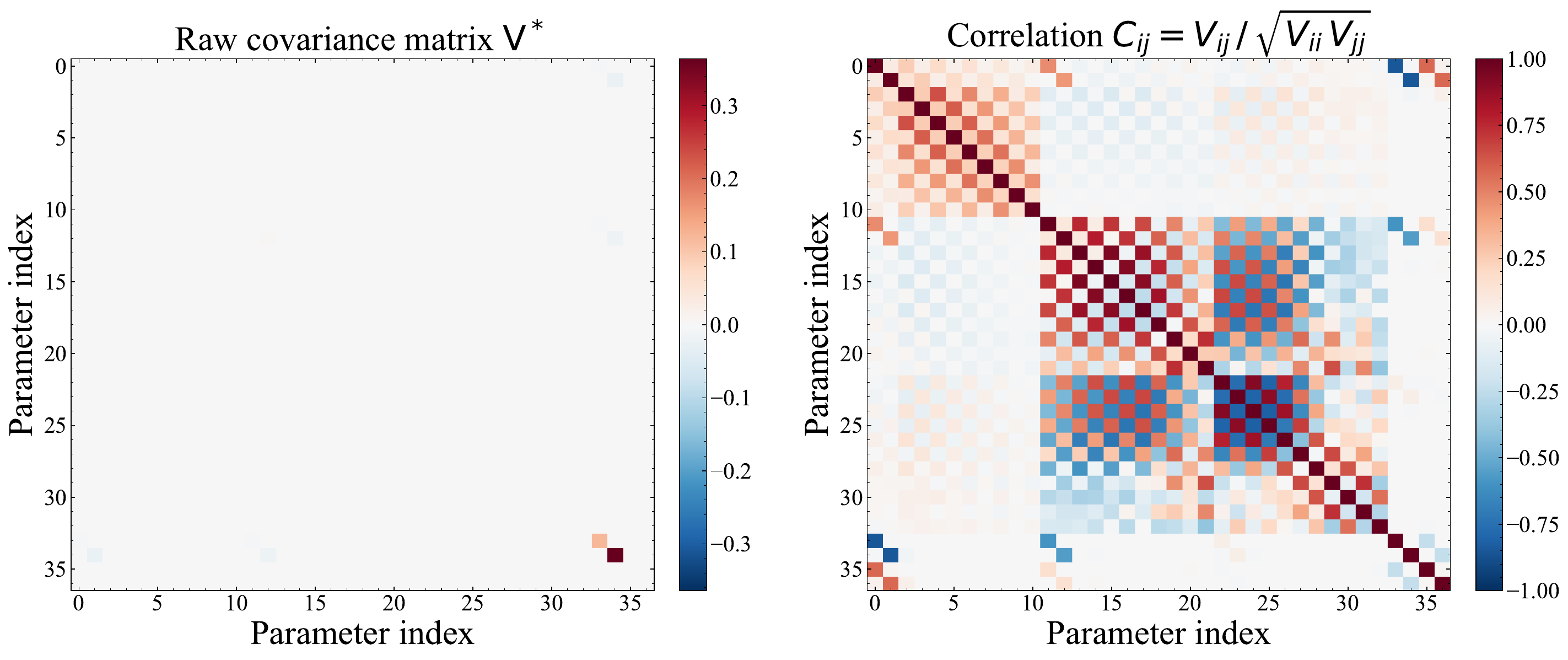}
    \caption{Posterior covariance matrix $\mathbf{V}^*$ (left) and correlation matrix $C_{ij}$ (right) for the Chebyshev pipeline at the same configuration as Figure~\ref{fig:corr_heatmap}. The intra-parameter correlation blocks now show the near-orthogonality of adjacent Chebyshev basis functions. This reduces $\kappa(\mathbf{V}^*)$ from $\sim10^{10}$ to $\sim10^4$.}
    \label{fig:corr_heatmap_cheby}
\end{figure*}

Figure~\ref{fig:corr_heatmap_cheby} shows the same covariance and correlation matrices for the Chebyshev pipeline at the same fixed orders $\mathbf{n} = [10,10,10,1,1]$, providing a direct comparison with Figure~\ref{fig:corr_heatmap}. The posterior covariance matrix of the new Chebyshev formalism shows an entirely different correlation structure. The solid bright blocks of the monomial case are replaced by an alternating red and blue checkerboard pattern within each NWP block. This demonstrates the orthogonality of the Chebyshev basis $T_j$ and $T_{j+1}$ oscillate with opposite parity over the band, so their coefficients anti-correlate. The result is a weaker and less coherent correlation structure, which is reflected in the reduction of $\kappa(\mathbf{V}^*)$ from $\sim10^{11}$ to $\sim10^4$. Nevertheless, residual correlations persist and a further source of ill-conditioning remains to be addressed.

\subsubsection{4-NWP pipeline}

Given the degeneracy between $\mathbf{X_\mathrm{L}}$ and $\mathbf{X}_\mathrm{NS}$ 
identified in Section~\ref{subsec:Degeneracy_in_X}, we propose a mitigation strategy that
fixes $T_\mathrm{NS}$ to its independently known value and fits only
the remaining four NWPs. Of the two degenerate parameters,
we choose to fix $T_\mathrm{NS}$ because it is
independently measurable: the excess noise ratio (ENR) of the noise 
diode used, NC346A, is provided by the manufacturer's datasheet, giving 
$T_\mathrm{NS} \approx 1100~\mathrm{K}$ \citep{noisecom_nc346}. We have also observed that fixing $T_\mathrm {NS}$ instead of $T_{\rm L}$ gives us a better condition number for the posterior covariance matrix and better evidence for the model.

This scalar 4-NWP model is numerically stable, but it assumes that the 
manufacturer-supplied ENR value is both accurate and frequency independent.
A data-driven frequency-dependent alternative based on the hot-load measurement 
is presented in Section~\ref{sec:hotload_iterative_tns}.

Recall from equation~\eqref{eq:linear_model} that $T_\mathrm{cal} = \sum_i X_i\, T_i + \sigma_\mathrm{cal}$.
Since $T_\mathrm{NS}$ is now fixed, its contribution can be computed
directly from the data and subtracted from the left-hand side of the equation. We define
the corrected calibration temperature
\begin{equation}
    \widetilde{T}_\mathrm{cal} \equiv T_\mathrm{cal} - X_\mathrm{NS}\, T_\mathrm{NS}^\mathrm{ENR},
    \label{eq:tcal_corrected}
\end{equation}
where $T_\mathrm{NS}^\mathrm{ENR} = 1100~\mathrm{K}$ is the fixed noise
source temperature. This yields a reduced linear system with only four
unknowns:
\begin{equation}
    \widetilde{T}_\mathrm{cal} = X_\mathrm{unc}\, T_\mathrm{unc} + X_\mathrm{cos}\, T_\mathrm{cos} + X_\mathrm{sin}\, T_\mathrm{sin} + X_\mathrm{L}\, T_\mathrm{L} + \sigma_\mathrm{cal},
\end{equation}
or equivalently $\widetilde{T}_\mathrm{cal} = \widetilde{\mathbf{X}} \widetilde{\boldsymbol{\Theta}} + \sigma_\mathrm{cal}$,
where $\widetilde{\boldsymbol{\Theta}} = (T_\mathrm{unc}, T_\mathrm{cos}, T_\mathrm{sin}, T_\mathrm{L})$
and $\widetilde{\mathbf{X}}$ is the design matrix with the $X_\mathrm{NS}$ column
removed. The Chebyshev pipeline described in
Section~\ref{sec:Chebyshev_pipeline} is then applied to this reduced system
to obtain the posterior for the four remaining noise wave parameters\footnote{For more discussion on the ENR and the degeneracies in the equation please refer to \cite{kirkham26}}.

By removing the $\mathbf{X}_\mathrm{NS}$ column from the design matrix, the
near-collinearity between $\mathbf{X}_\mathrm{NS}$ and $\mathbf{X}_\mathrm{L}$ is eliminated
entirely, since $T_\mathrm{NS}$ is no longer a free parameter. This
directly addresses the root cause of the ill-conditioning identified
above. Hereafter we refer to this approach as the `4-NWP' fit, while the
Chebyshev pipeline approach with all five NWPs is referred to as the `5-NWP' fit.

\begin{figure}
    \centering
    \includegraphics[width=0.48\textwidth]{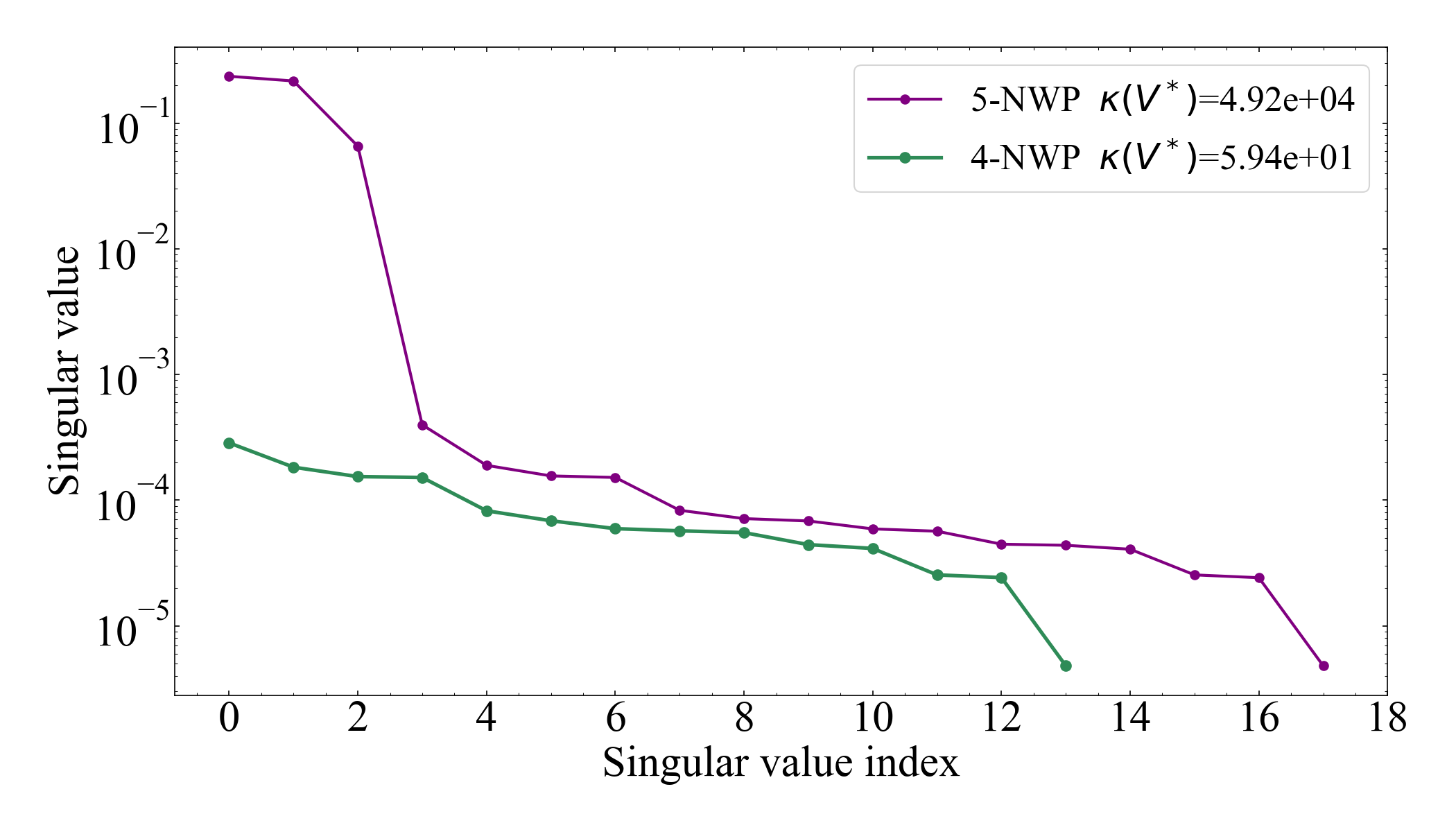}
    \caption{SVD spectra of the posterior $\mathbf{V}^*$ for the 5-NWP fit
    (purple, $\kappa\approx4.9\times10^{4}$) and the 4-NWP two-step fit with
    $T_\mathrm{NS}$ fixed to $1100$~K (green, $\kappa\approx5.94\times10^{1}$).
    By fixing $T_\mathrm{NS}$, the 4-NWP model has fewer free parameters, 
    so its SVD spectrum contains fewer singular values. The parameters associated 
    with the $X_\mathrm{NS}$/$X_\mathrm{L}$ degeneracy no longer appear, removing 
    the largest singular values from the spectrum and reducing the condition number
    by nearly three orders of magnitude.}
    \label{fig:4nwp_svd}
\end{figure}

Figure~\ref{fig:4nwp_svd} compares the SVD of $\mathbf{V}^*$ between the 5-NWP fit and the 4-NWP two-step fit on the mock data.
The condition number drops from $\kappa(\mathbf{V}^*) \approx 4.9\times10^{4}$ for the 5-NWP fit to $\kappa(\mathbf{V}^*) \approx 5.9\times10^{1}$ for the 4-NWP fit, which is a reduction of nearly three orders of magnitude.

\subsubsection{Iterative hot-load recovery of $T_\mathrm{NS}(\nu)$}
\label{sec:hotload_iterative_tns}

The scalar 4-NWP model removes the dominant $X_\mathrm{NS}/X_\mathrm{L}$ degeneracy by taking
$T_\mathrm{NS}$ out of the fit. This is numerically effective, but it is only exact if the
noise diode is both known and spectrally flat. In the mock and real REACH data the hot
calibration load provides a more direct constraint: its physical temperature is known and
its noise-source coupling ($\rm X_{NS}$) is large compared to the cold load, so the residual contribution
of the noise diode can be isolated after subtracting the other four noise-wave terms.

This use of absolute load measurements to constrain the calibration parameters is closely
related to the iterative EDGES receiver calibration of \citet{monslave17}. Their
formulation describes the internal-mismatch degrees of freedom with frequency-dependent
scale and offset corrections, $C_1(\nu)$ and $C_2(\nu)$, which correct the initially
assumed $T_\mathrm{NS}$ and $T_\mathrm{L}$ and also absorb small reference-plane and
mismatch effects. EDGES alternates updates of $C_1$ and $C_2$ from ambient- and hot-load
measurements with fits of $T_\mathrm{unc}$, $T_\mathrm{cos}$, and $T_\mathrm{sin}$ from
open- and short-cable measurements until convergence. Our method follows the same broad
self-consistency principle, but parameterises the internal mismatch and
reconstructs $T_\mathrm{NS}(\nu)$ from the hot-load residual while retaining
$T_\mathrm{L}(\nu)$ as the fourth fitted NWP. \citet{murray22} and
\citet{murray26} make the correspondence between the two parameterisations
explicit. $C_1$ and $C_2$ are a scale and an offset applied to initial guesses of
the internal-load temperatures, $T'_\mathrm{NS} = C_1 T_\mathrm{NS}^\mathrm{guess}$
and $T'_\mathrm{L} = T_\mathrm{L}^\mathrm{guess} - C_2$, and the guesses only affect
the convergence rate of the iteration \citep{murray26}. \citet{murray22} fit these
internal-load temperatures directly within a Bayesian calibration likelihood, and
\citet{murray26} describe the iterative loop in terms of $T'_\mathrm{NS}$ and
$T'_\mathrm{L}$ themselves. The iterative approach therefore does not depend on the
$C_1/C_2$ parameterisation and can be applied directly to $T_\mathrm{NS}$ and
$T_\mathrm{L}$, as we do here.

We therefore resort to an iterative
hot-load measurement to calculate $T_{\rm NS}(\nu)$. Starting from a scalar guess $T_\mathrm{NS}^{(0)}$, we run the 4-NWP fit and
recover the four remaining parameters. The hot-load equation is then rearranged at each
frequency channel to give
	\begin{align}
	    T_\mathrm{NS}^{(k+1)}(\nu) &=
	    \frac{R_\mathrm{hot}^{(k)}(\nu)}{X_\mathrm{NS}^\mathrm{hot}(\nu)},
    \label{eq:hotload_tns_update}\\
    \rm where,\,
    R_\mathrm{hot}^{(k)}(\nu) &= T_\mathrm{cal}^\mathrm{hot}(\nu)
    - X_\mathrm{unc}^\mathrm{hot}T_\mathrm{unc}^{(k)}
    - X_\mathrm{cos}^\mathrm{hot}T_\mathrm{cos}^{(k)} \nonumber\\
    &\quad
	    - X_\mathrm{sin}^\mathrm{hot}T_\mathrm{sin}^{(k)}
	    - X_\mathrm{L}^\mathrm{hot}T_\mathrm{L}^{(k)},
	\end{align}
	Here $k$ denotes the iteration number, $T_\mathrm{cal}^\mathrm{hot}(\nu)$ is the
	measured calibrated temperature of the hot-load observation, and
	$X_i^\mathrm{hot}(\nu)$ are the hot-load design-matrix weights for the
	uncorrelated, cosine, sine, noise-source, and load terms. The quantities
	$T_\mathrm{unc}^{(k)}(\nu)$, $T_\mathrm{cos}^{(k)}(\nu)$,
	$T_\mathrm{sin}^{(k)}(\nu)$, and $T_\mathrm{L}^{(k)}(\nu)$ are the four NWP
	curves recovered by the 4-NWP fit at iteration $k$. The residual
	$R_\mathrm{hot}^{(k)}(\nu)$ is therefore the part of the hot-load temperature
	left after subtracting these four fitted contributions; dividing by
	$X_\mathrm{NS}^\mathrm{hot}(\nu)$ converts this residual into the next
	estimate of the noise-source temperature.
	The raw channel-by-channel estimate is noisy because division by the hot-load
	$X_\mathrm{NS}$ response also amplifies channel-scale measurement noise. We therefore
	regularise each update by projecting $T_\mathrm{NS}^{(k+1)}(\nu)$ onto a second-order
	Chebyshev polynomial in normalised physical frequency before feeding it back into the
	next 4-NWP fit. This retains the broadband level, slope, and curvature expected from the
	noise diode and its analogue signal path, while preventing unsupported channel-scale
	structure from propagating through the iteration. Frequency binning could likewise
	reduce the variance, but would introduce choices of bin width and edges and produce a
	piecewise estimate that must be interpolated back onto the full calibration grid. The
	low-order polynomial instead uses all valid channels to provide a continuous update with
	only three degrees of freedom. The iteration is stopped when the rms and maximum change
	between successive smoothed $T_\mathrm{NS}(\nu)$ curves fall below fixed tolerances.

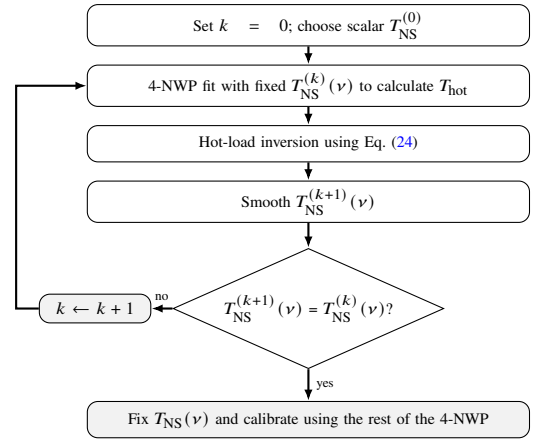
\begin{figure}
	    \centering
	    \begin{tikzpicture}[
	        node distance=0.24cm,
	        proc/.style={draw, rounded corners, align=center, text width=0.66\columnwidth, minimum height=0.48cm, font=\scriptsize},
	        decision/.style={draw, diamond, aspect=2.25, align=center, inner sep=0.5pt, text width=0.31\columnwidth, font=\scriptsize},
	        increment/.style={draw, rounded corners, align=center, text width=0.15\columnwidth, minimum height=0.38cm, font=\scriptsize, fill=black!5},
	        finalbox/.style={draw, rounded corners, align=center, text width=0.66\columnwidth, minimum height=0.48cm, font=\scriptsize, fill=black!5},
	        arrow/.style={-{Latex[length=1.5mm]}, thick}
	    ]
	        \node[proc] (guess) {Set $k=0$; choose scalar $T_\mathrm{NS}^{(0)}$};
	        \node[proc, below=of guess] (fit) {4-NWP fit with fixed $T_\mathrm{NS}^{(k)}(\nu)$ to calculate $T_{\rm hot}$};
	        \node[proc, below=of fit] (hot) {Hot-load inversion using Eq.~\eqref{eq:hotload_tns_update}};
	        \node[proc, below=of hot] (smooth) {Smooth $T_\mathrm{NS}^{(k+1)}(\nu)$};
	        \node[decision, below=0.33cm of smooth] (conv) {$T_\mathrm{NS}^{(k+1)}(\nu)$ = $T_\mathrm{NS}^{(k)}(\nu)$?};
	        \node[increment, left=0.28cm of conv] (inc) {$k\leftarrow k+1$};
	        \node[finalbox, below=0.42cm of conv] (final) {Fix $T_\mathrm{NS}(\nu)$ and calibrate using the rest of the 4-NWP};
	        \draw[arrow] (guess) -- (fit);
	        \draw[arrow] (fit) -- (hot);
	        \draw[arrow] (hot) -- (smooth);
	        \draw[arrow] (smooth) -- (conv);
	        \draw[arrow] (conv) -- node[right, font=\tiny] {yes} (final);
	        \draw[arrow] (conv.west) -- node[above, font=\tiny] {no} (inc.east);
	        \draw[arrow] (inc.west) -- ++(-0.30,0) |- (fit.west);
	    \end{tikzpicture}
	    \caption{A representative flowchart 
        for the iterative hot-load $T_\mathrm{NS}(\nu)$ recovery. Here, $k$ denotes the interation number.
        Each loop fits only the four non-degenerate NWPs while holding 
        $T_\mathrm{NS}(\nu)$ fixed, then updates the fixed curve from the hot-load residual. 
        The iteration stops once the smoothed $T_\mathrm{NS}(\nu)$ update has converged.}
	    \label{fig:hotload_tns_flowchart}
	\end{figure}

This approach keeps the conditioning advantage of the 4-NWP model because
$T_\mathrm{NS}(\nu)$ is fixed during each individual fit; the $X_\mathrm{NS}$ column is not
reintroduced as an unconstrained polynomial parameter. At the same time, it removes the
most important limitation of the scalar 4-NWP model by allowing the noise-source
temperature to vary smoothly across the band. The iterative hot-load $\rm T_{NS}(\nu)$ measurement calibration process is explained using a flow-chart in Figure \ref{fig:hotload_tns_flowchart}.
Within each iteration, the polynomial orders of the four fitted NWPs are still
selected by maximising the Bayesian evidence of the 4-NWP fit. Only the degree of the
$T_\mathrm{NS}(\nu)$ smoothing polynomial is fixed, because the hot-load inversion lies
outside the likelihood.

An alternative to the iteration would be to keep $T_\mathrm{NS}$ in a 5-NWP fit
and constrain its polynomial coefficients with informative priors. We tested one such
prior family on the mock data, with c2r91 withheld and all polynomial orders re-selected
by the evidence for each prior width. The constant coefficient has prior mean
$1100.49$~K and the higher-order coefficients have zero prior means. We scale the
variances of all $T_\mathrm{NS}$ coefficients together, retaining the existing
degree-dependent weighting. A constant prior mean therefore still permits
frequency-dependent posterior structure at finite prior width. Widths quoted below
are conditional prior standard deviations of the constant coefficient evaluated at
the fitted noise level. At $\sim10$~K, an illustrative per-cent-level ENR uncertainty,
$\kappa(\mathbf{V}^*)\approx5\times10^{4}$, similar to the Chebyshev 5-NWP fit.
At $\sim0.1$~K, the selected $T_\mathrm{NS}$ order is five, the antenna RMSE improves
from $0.310$ to $0.268$~K, and $\kappa(\mathbf{V}^*)$ decreases to $1.9\times10^{4}$.
Tightening the prior further suppresses the higher-order coefficients as well as
constraining the constant term. At $\sim0.01$ and $\sim0.001$~K,
$\kappa(\mathbf{V}^*)$ falls to approximately $1.2\times10^{3}$ and $90$, respectively,
while the antenna RMSE rises to $0.86$--$0.94$~K, approaching the scalar 4-NWP result.
Thus this prior family yields a tradeoff between conditioning and calibration accuracy.
These tests leave open other choices of coefficient covariance or frequency-dependent
prior mean. The iterative hot-load method provides one way to estimate a smooth
$T_\mathrm{NS}(\nu)$ curve from the data while keeping it fixed within each 4-NWP fit.

\subsection{Cable standing-wave degeneracies: a third source of instability}
\label{sec:cable_masking}

The Chebyshev basis and the 4-NWP approach together address the two sources of 
ill-conditioning identified above. A third, independent source of instability may arise in any global 
signal experiment that employs cable-connected calibrators of finite electrical length. 
Standing-wave degeneracies in the cable $S_{11}$ produce periodic oscillations in the $X$ 
parameters as a function of frequency. At some frequencies, calibrators $|S_{11}|$ values can make the corresponding $\mathbf{X}$-matrix columns near-collinear 
and produce localised spikes in $\kappa(\mathbf{X})$. 
Unlike the sources addressed above, this instability depends on the specific calibrator and cable
choice and cannot be mitigated by the Chebyshev basis or the 4-NWP fix alone.

A closely related instability has recently been reported for EDGES by
\citet{murray26}. In their iterative calibration scheme the internal-load
temperatures are fitted to the ambient and hot loads separately from the
noise-wave terms, so the ill-conditioning they find cannot arise from the
$X_\mathrm{NS}/X_\mathrm{L}$ degeneracy of Section~\ref{subsec:Degeneracy_in_X}.
Instead, the condition number of their noise-wave fit rises from
$\sim4.5\times10^{4}$ to $\sim7.5\times10^{5}$ as the fitted band narrows from
$50$--$100$ to $60$--$90$~MHz. They attribute this to the limited frequency
structure of the cable reflections: $T_\mathrm{unc}$, $T_\mathrm{cos}$ and
$T_\mathrm{sin}$ share the same basis functions and are separated only through
their products with reflection-dependent coefficients, so when the cable
reflections have too few turning points per model coefficient within the band,
$T_\mathrm{cos}$ and $T_\mathrm{sin}$ become strongly correlated. They therefore
recommend at least one turning point per model coefficient. This is the same
physical mechanism as the cable standing-wave degeneracy studied here, because
$X_\mathrm{cos}$ and $X_\mathrm{sin}$ (equations~\ref{eq:x_cos}
and~\ref{eq:x_sin}) inherit their frequency structure from the cable reflection
coefficients. The two analyses view it from complementary directions.
\citet{murray26} consider the global polynomial fit and ask whether the cable
reflections oscillate enough across the band to separate the polynomial
coefficients, while we examine the design matrix channel by channel and locate
the individual frequencies at which the calibrator rows become near-collinear.
Their turning-point criterion is also consistent with Section~\ref{sec:diagnostics},
where the highest-order polynomial coefficients are the directions that the
calibrator data cannot constrain.

We investigate this effect over the $90$--$130$~MHz subband rather than the full $50$--$130$~MHz band.
This choice is motivated by the fact that it places the strongest condition number spikes close to the band edge. This would present a robust test case for the exercise described in this section, for two primary reasons.
First, even before any masking is applied, a collinearity spike at the edge corrupts the global
Chebyshev polynomial fit more severely than an equivalent interior spike. Because the calibration
uses a global polynomial fit over all frequencies simultaneously, the ill-conditioned design-matrix
rows at the spike frequency contaminate the posterior through $\mathbf{X}^\top\mathbf{X}$.
After normalising the frequency to $[-1,1]$, every Chebyshev basis function satisfies 
$|T_n(\pm1)|=1$ at the edges while oscillating below unity in the interior \citep{trefethen13}. 
A spike at the band edge therefore, biases the polynomial fit more
strongly than an equivalent interior spike, because all the remaining calibrated channels lie
on only one side and cannot pull the solution back. An interior spike, by contrast,
is supported by the data on both sides of the band, which constrains the fit and limits the instability.
Second, if the spike is subsequently masked, the edge location makes the remedy more costly.
The statistical leverage of a frequency channel,
$h(\nu) = \mathbf{g}(\nu)^\top(\mathbf{X}^\top\mathbf{X})^{-1}\mathbf{g}(\nu)$
\citep{belsey05}, measures how strongly that channel pulls the final fit, where
$\mathbf{g}(\nu) = [T_0(\nu_*),\, T_1(\nu_*),\, \ldots,\, T_n(\nu_*)]$ is the vector of
Chebyshev basis functions evaluated at the normalised frequency $\nu_* \in [-1,1]$.
Leverage is determined by the predictor value $\nu_*$. Channels at the band edge
have the most extreme predictor values ($\nu_* \to \pm 1$), at which every Chebyshev
basis function simultaneously reaches $|T_k(\pm1)| = 1$ \citep{trefethen13}, maximising
$\|\mathbf{g}\|$ and therefore $h(\nu)$.
Removing such a high-leverage channel leaves the polynomial unconstrained on one side
and the fit variance diverges under extrapolation beyond the masked region.
We choose the subband $90$--$130$~MHz to shift the strongest of the condition number peaks
($\kappa(\mathbf{X}) \approx 3.9\times10^{3}$ at $94.8$~MHz) toward the lower edge,
providing a more stringent test of both the unmasked instability and the masking procedure.
Other than that, \cite{monslave17} also suggest that a subband calibration may be helpful to 
optimize calibration performance, which we will explore in a future work.

\begin{figure}
    \centering
    \includegraphics[width=0.48\textwidth]{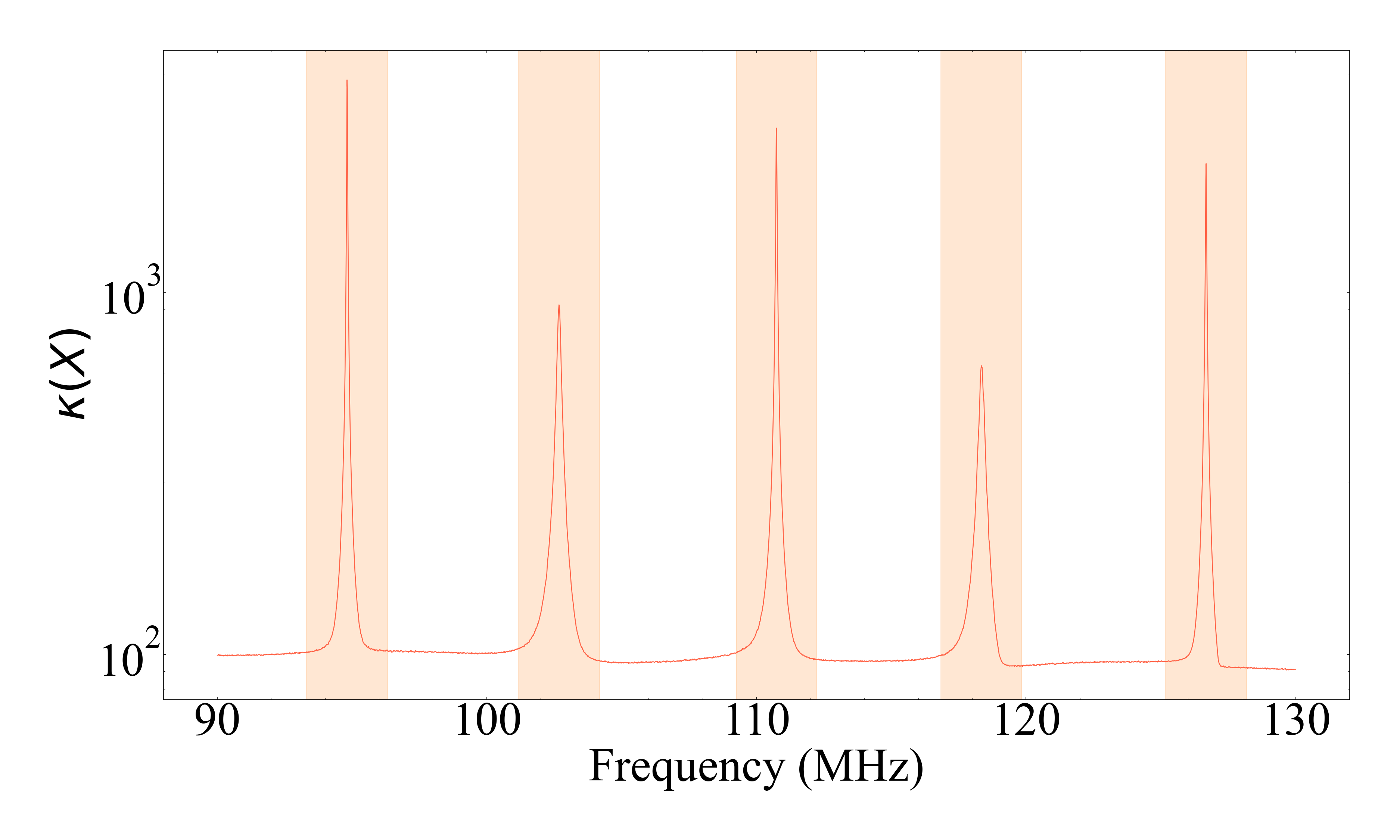}
    \caption{Condition number $\kappa(\mathbf{X})$ of the calibration design matrix 
    over $90$--$130$~MHz for the calibrator set 
    \texttt{hot}, \texttt{c2r27}, \texttt{c2r36}, \texttt{c2r69}, \texttt{c10short}, \texttt{c10r10}. 
    Localised spikes from 10-metre cable standing-wave collinearities appear with 
    $\approx 6.3$~MHz spacing, with additional broader features from the 2-metre cables. 
    Orange shaded bands show the masked windows ($\pm1.5$~MHz around each detected spike) 
    used in cases~(c) and~(d) described in Section~\ref{sec:cable_masking}.}
    \label{fig:bad_cal_cn}
\end{figure}

To investigate this effect, we select a calibrator subset containing: 
\texttt{hot}, \texttt{c2r27}, \texttt{c2r36}, \texttt{c2r69}, \texttt{c10short}, 
and \texttt{c10r10} with a mix of 2-metre and 10-metre cables. 
The 10-metre cables introduce standing-wave collinearities spaced by 
$\Delta\nu \approx 6.3$~MHz across the $90$--$130$~MHz band; 
the 2-metre cables add collinearities with broader spacing. 
Figure~\ref{fig:bad_cal_cn} shows the per-frequency condition number 
$\kappa(\mathbf{X})$ for this calibrator set. 
Pronounced spikes arise at certain frequencies where the $X_i$ terms become locally near-collinear. 
The orange shaded bands indicate frequency channels masked at 
$\pm1.5$~MHz around each detected spike.

We compare four calibration strategies on this calibrator set:
\begin{enumerate}
    \item[(a)] 5-NWP, unmasked: full fit including $T_\mathrm{NS}$, 
    affected by both the $X_\mathrm{NS}/X_\mathrm{L}$ degeneracy and 
    cable collinearity spikes ($\kappa \approx 3\times10^4$);
    \item[(b)] 4-NWP fixed $T_\mathrm{NS}$, unmasked: the degeneracy is 
    removed but cable induced spikes persist ($\kappa \approx 6.9\times10^3$);
    \item[(c)] 4-NWP fixed $T_\mathrm{NS}$, masked: spike channels masked 
    at $\pm1.5$~MHz around each detected peak, reducing $\kappa$ to $\approx 1.3\times10^3$;
    \item[(d)] 4-NWP hot-load $T_\mathrm{NS}(\nu)$, masked: the
    frequency-dependent $T_\mathrm{NS}(\nu)$ curve is recovered from the hot-load data
    using the iterative method of Section~\ref{sec:hotload_iterative_tns}, and the
    final 4-NWP calibration is run on the masked calibrator data.
\end{enumerate}
In Section~\ref{sec:results} we show the effect of each strategy on the calibration residuals.
\section{Results}
\label{sec:results}

In this section, we show the calibrated temperatures for the validator source c2r91 and the antenna
for each of the four calibration strategies.

Table~\ref{tab:pipeline_comparison} summarises the four main pipeline 
configurations tested on the mock dataset over the full $50$--$130$~MHz band. 
The masking test of Section~\ref{sec:cable_masking} uses the 
$90$--$130$~MHz subband.
To compare the residuals with the noise floor of the mock data, we also compute the
pipeline's expected measurement-noise RMSE from the measured Dicke-switch spectra. This
estimate follows the noise-band calculation implemented in the calibration plotting code,
which estimates the PSD fluctuations empirically. For each source, we smooth
the three measured power spectra $P_\mathrm{cal}$, $P_\mathrm{L}$, and $P_\mathrm{NS}$
with a $0.5$~MHz Savitzky--Golay filter and use the residuals about these smooth curves
to estimate the per-channel power fluctuations $\sigma_{P_\mathrm{cal}}$,
$\sigma_{P_\mathrm{L}}$, and $\sigma_{P_\mathrm{NS}}$ and their covariances. Propagating
these fluctuations through the Dicke calibration equation gives
\begin{align}
    \sigma_T(\nu) &=
    \frac{T_\mathrm{NS}(\nu)X_\mathrm{L}(\nu)}
    {P_\mathrm{NS}(\nu)-P_\mathrm{L}(\nu)}
    \left[V_P(\nu)\right]^{1/2},
    \label{eq:thermal_rmse_noise}\\
    V_P &=
    |\sigma_{P_\mathrm{cal}}^2+\sigma_{P_\mathrm{L}}^2
    -2\,\mathrm{Cov}(P_\mathrm{cal},P_\mathrm{L}) \nonumber\\
    &\quad
    + Q^2\{\sigma_{P_\mathrm{NS}}^2+\sigma_{P_\mathrm{L}}^2
    -2\,\mathrm{Cov}(P_\mathrm{NS},P_\mathrm{L})\}
    -2Q\,\mathrm{Cov}(A,B)|.
\end{align}
where $Q=(P_\mathrm{cal}-P_\mathrm{L})/(P_\mathrm{NS}-P_\mathrm{L})$,
$A=P_\mathrm{cal}-P_\mathrm{L}$, and $B=P_\mathrm{NS}-P_\mathrm{L}$ \citep{kirkham25}.
After applying the same source-plane temperature correction used for the residuals, the
expected-noise RMSE reported in Table~\ref{tab:pipeline_comparison} is
\begin{equation}
    \mathrm{RMSE}_\mathrm{noise} =
    \left[\frac{1}{N_\nu}\sum_\nu \sigma_T^2(\nu)\right]^{1/2},
\end{equation}
computed over $\rm N_\nu$ number of unbinned channels in the same frequency range.

\begin{table*}
    \centering
    \footnotesize
    \setlength{\tabcolsep}{3.5pt}
    \caption{Progressive stabilisation of the calibration pipeline on the mock dataset over the
    full $50$--$130$~MHz band. Each row introduces one modification to the previous configuration.
    $\kappa(\mathbf{V}^*)$ is the condition number of the posterior covariance matrix.
    Antenna and validator (c2r91) RMSEs are unbinned channel-space values computed against the injected mock truth.
    The noise RMSE column gives the expected antenna/validator channel-space measurement-noise floor,
    computed by propagating the measured Dicke-switch PSD fluctuations over the same frequency mask.
    $\ln\mathcal{Z}$ is evaluated independently under NumPy~1.x and NumPy~2.x backends;
    agreement after the Chebyshev and 4-NWP steps confirms numerical reproducibility.}
    \label{tab:pipeline_comparison}
        \begin{tabular}{lcccccc}
        \hline
        Configuration & $\kappa(\mathbf{V}^*)$
            & $\ln\mathcal{Z}_{1.x}$ & $\ln\mathcal{Z}_{2.x}$
            & Ant.\ RMSE (K) & Val.\ RMSE (K)
            & Noise RMSE Ant./Val. (K) \\
        \hline
        Monomial 5-NWP  & $4.42\times10^{11}$ & $-50934.01$ & $-125980.12$ & $0.309$ & $0.092$ & $0.262/0.096$ \\
        Chebyshev 5-NWP & $4.92\times10^{4}$  & $50418.65$  & $50418.65$  & $0.314$ & $0.092$ & $0.262/0.096$ \\
        Chebyshev 4-NWP & $5.94\times10^{1}$  & $50236.84$  & $50236.84$  & $0.937$ & $0.092$ & $0.262/0.096$ \\
        4-NWP hot-load $T_\mathrm{NS}(\nu)$ & $5.94\times10^{1}$ & $44718.50$ & $44718.50$ & $0.264$ & $0.092$ & $0.262/0.096$ \\
        \hline
    \end{tabular}
\end{table*}

We validate the stabilised pipeline on the mock data over the full
$50$--$130$~MHz band. A scalar 4-NWP run with $T_\mathrm{NS} \approx 1100$~K is already
well-conditioned, but it leaves a broad chromatic residual in the antenna prediction
because the injected noise-source temperature is not perfectly flat. The iterative
hot-load method of Section~\ref{sec:hotload_iterative_tns} uses the hot-load spectrum to
recover this frequency dependence directly.

\begin{figure}
    \centering
    \includegraphics[width=\columnwidth]{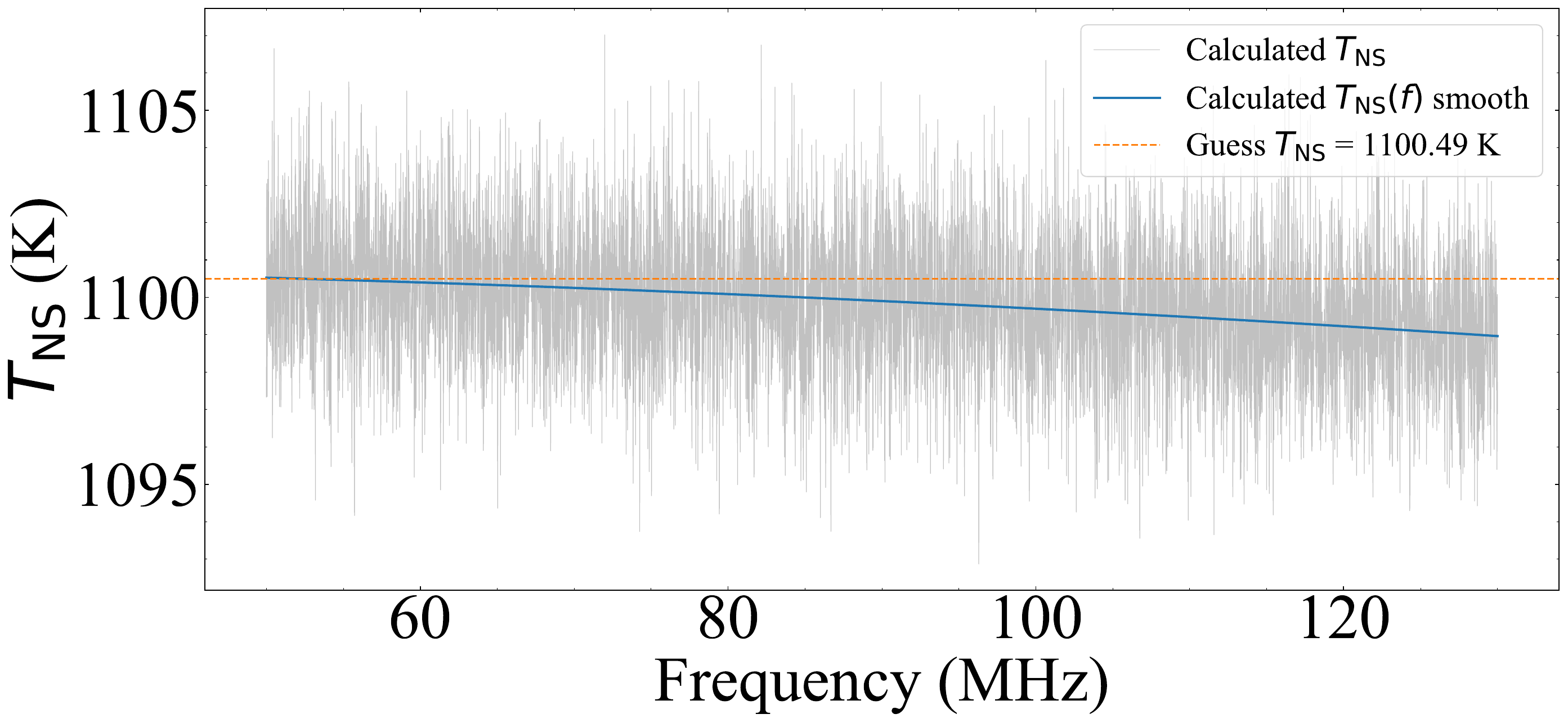}
    \caption{Hot-load recovery of the frequency-dependent noise-source temperature for the
    mock dataset. Grey shows the raw channel-by-channel hot-load inversion,
    blue shows the smoothed $T_\mathrm{NS}(\nu)$ curve used in the 4-NWP calibration, and
    the dashed orange line marks the scalar guess $T_\mathrm{NS}=1100.49$~K.}
    \label{fig:hotload_tns_recovery}
\end{figure}

Figure~\ref{fig:hotload_tns_recovery} shows the recovered $T_\mathrm{NS}(\nu)$ curve. The
raw inversion is noisy at the channel level, but its smoothed component is stable and
captures the shallow spectral slope that is absent from the scalar ENR approximation.
This recovered curve is then held fixed in the final 4-NWP calibration, preserving the
conditioning improvement of the reduced model.

\begin{figure*}
    \centering
    \includegraphics[width=\textwidth]{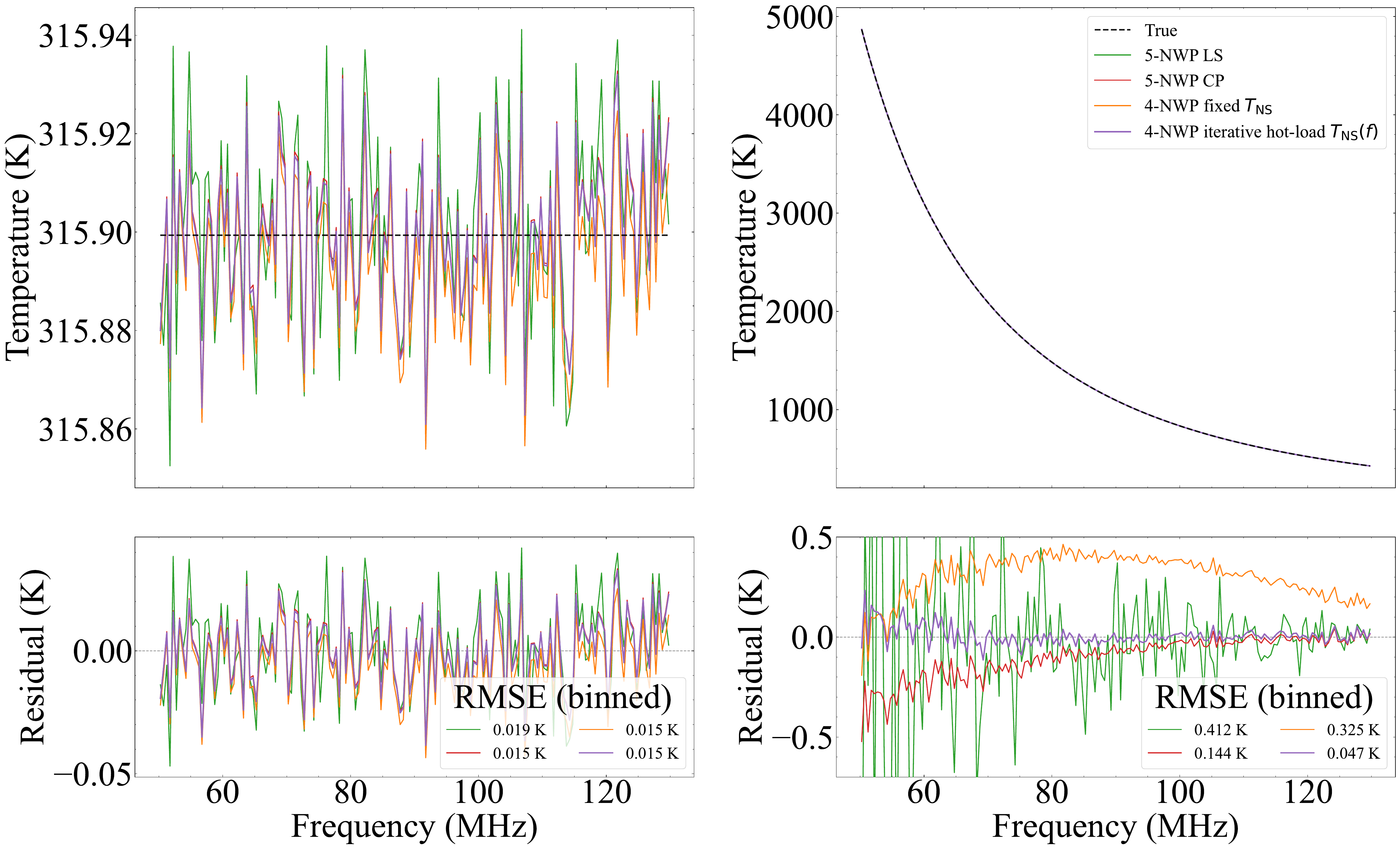}
    \caption{Four-method comparison on the mock dataset. The panels show the
    validator c2r91 and antenna temperatures, with residuals below each temperature panel.
    The iterative hot-load 4-NWP method uses the recovered $T_\mathrm{NS}(\nu)$ curve from
    Fig.~\ref{fig:hotload_tns_recovery}. It retains sub-noise validator residuals and
    strongly reduces the broad antenna residual left by the fixed-scalar 4-NWP fit.}
    \label{fig:hotload_four_method}
\end{figure*}

Figure~\ref{fig:hotload_four_method} compares the five-parameter fit, the scalar 4-NWP
fit, and the iterative hot-load 4-NWP fit on the same mock dataset. Replacing the scalar with
the recovered hot-load $T_\mathrm{NS}(\nu)$ removes most of the chromatic structure left by the fixed 4-NWP fit: the
binned antenna RMSE falls to $0.047$~K, while the validator RMSE remains at $0.015$~K. The
hot-load update therefore keeps the numerical gain of the 4-NWP model while recovering the
spectral information that was lost when $T_\mathrm{NS}$ was fixed to a single number.

\begin{figure}
    \centering
    \includegraphics[width=\columnwidth]{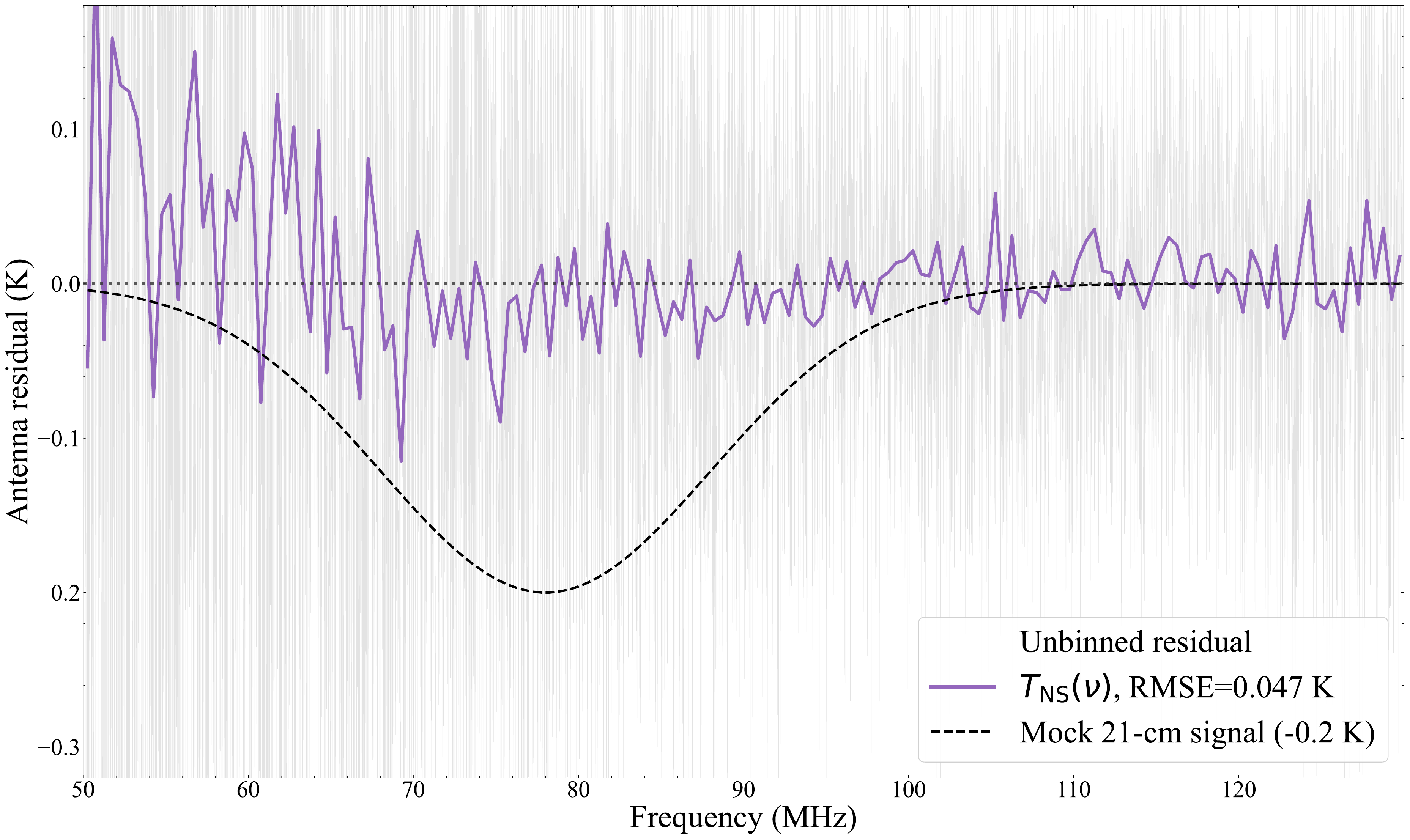}
    \caption{Antenna residual for the mock iterative hot-load
    $T_\mathrm{NS}(\nu)$ calibration over $50$--$130$~MHz. Grey shows the unbinned
    channel-space residual and purple shows the $0.5$~MHz binned residual, with a
    binned RMSE of $0.047$~K. The dashed black curve is a Gaussian-like mock
    global 21-cm absorption profile with amplitude $-0.2$~K, plotted only as a
    scale reference for the residuals.}
    \label{fig:hotload_ant_21cm_scale}
\end{figure}

The science requirement for global 21-cm experiments is not simply to minimise the
absolute calibration RMSE, but to suppress coherent spectral residuals below the scale of
the cosmological signal after foreground modelling. Figure~\ref{fig:hotload_ant_21cm_scale}
therefore isolates the antenna residual from the iterative hot-load solution and overlays
a Gaussian-like mock global 21-cm absorption profile with a $0.2$~K amplitude. This
profile is not injected into the mock data or used in the calibration; it is included only
as a visual scale reference. The binned residuals from the recovered
$T_\mathrm{NS}(\nu)$ calibration remain below this representative signal scale across most
of the band, showing that the hot-load route removes the broad scalar-$T_\mathrm{NS}$
calibration structure to a level relevant for subsequent 21-cm signal recovery.

The monomial 5-NWP pipeline produces a log-evidence that differs 
by $\sim75\,000$ between the two NumPy backends, yielding 
different calibration residuals over the $90-130$~MHz band.
This shows that the instability leaks primarily into the log-evidence, 
which controls the order selection of the NWP, ultimately affecting
the calibration residuals.
In contrast, both Chebyshev configurations produce bit-identical 
log-evidence values across backends, and the Chebyshev 4-NWP 
calibrated temperatures agree to within numerical limits between 
software environments six orders of magnitude better than the monomial case.

\begin{figure}
    \centering
    \includegraphics[width=0.48\textwidth]{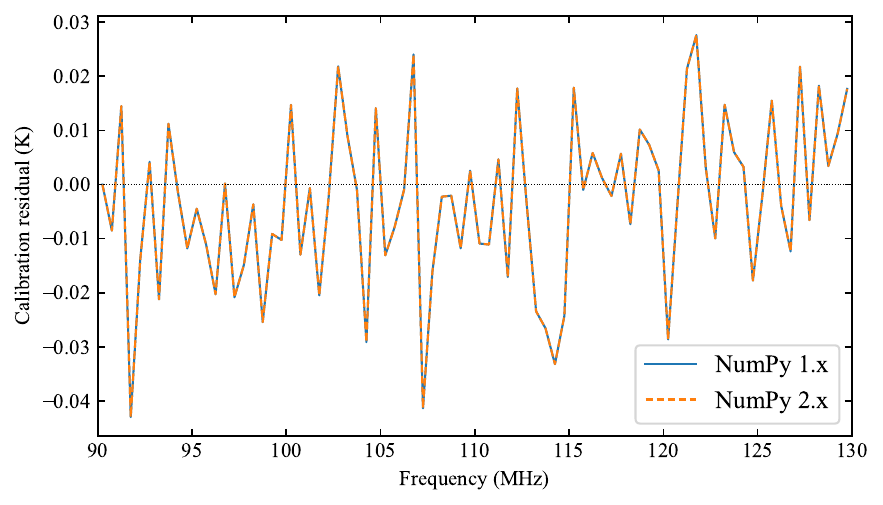}
    \caption{Calibration residuals (fitted $-$ true temperature) 
    for the validator source c2r91 under the Chebyshev 4-NWP pipeline, 
    binned to $0.5$~MHz, computed independently under NumPy~1.x 
    (\texttt{OpenBLAS}, blue) and NumPy~2.x (\texttt{Accelerate}, 
    orange dashed). The two curves are indistinguishable, showing 
    that the new 4-NWP pipeline is fully reproducible across software 
    environments. This is in contrast with Fig.~\ref{fig:residual_overlay}, 
    which shows the analogous plot for the monomial pipeline where 
    the two backends diverge by up to $\sim0.3~\mathrm{K}$.}
    \label{fig:reproducibility_restored}
\end{figure}

Figure~\ref{fig:reproducibility_restored} confirms this hypothesis -- 
the calibration residuals for the validator source (c2r91) are 
identical between NumPy~1.x and NumPy~2.x for the Chebyshev 4-NWP pipeline, 
in contrast to the divergence seen in Fig.~\ref{fig:residual_overlay} for 
the monomial pipeline.

\begin{figure*}
    \centering
    \includegraphics[width=\textwidth]{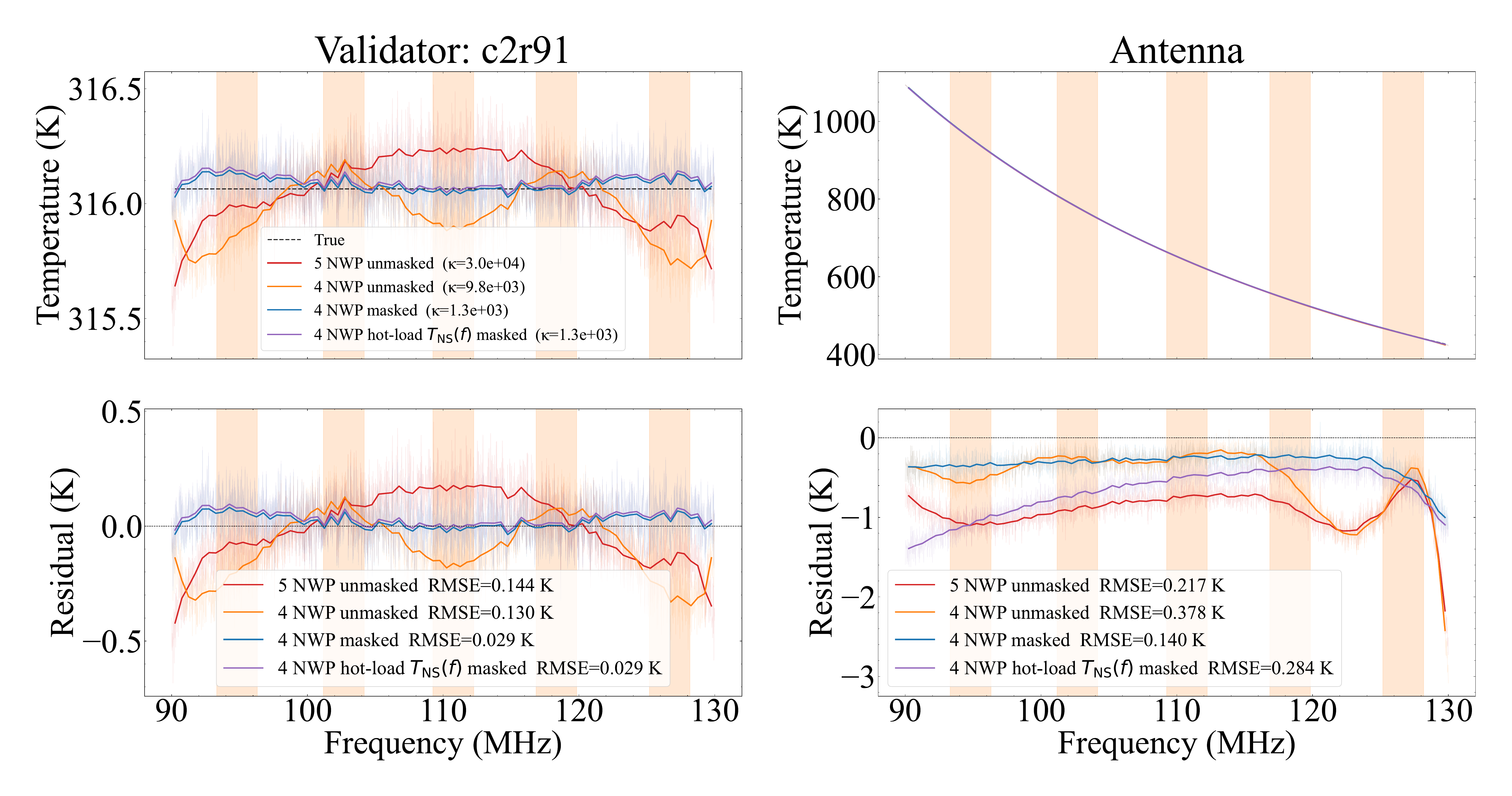}
    \caption{Calibrated validator (c2r91, left) and antenna (right) temperatures over $90$--$130$~MHz for four calibration strategies on the bad calibrator set. \textit{Top:} calibrated temperature; dashed black line shows the true mock spectrum. \textit{Bottom:} residuals (fitted $-$ true), binned to $0.5$~MHz with unbinned data in the background. Orange shaded bands indicate masked frequency channels. Case~(d) keeps the masked 4-NWP conditioning advantage but gives a larger antenna RMSE because the hot-load $T_\mathrm{NS}(\nu)$ recovery absorbs residual bad-calibrator and masked-edge modelling errors, which are amplified when propagated to the antenna.}
    \label{fig:bad_cal_comparison}
\end{figure*}

Figure~\ref{fig:bad_cal_comparison} shows the calibrated validator 
(c2r91) and antenna temperatures for the four cases of 
Section~\ref{sec:cable_masking}. Moving from case~(a) to case~(b), 
fixing $T_\mathrm{NS}$ breaks the $X_\mathrm{NS}/X_\mathrm{L}$ degeneracy 
but cable induced spikes leave structured residuals in the validator. 
Masking in case~(c) removes these artefacts, yielding validator and 
antenna RMSEs of $\sim0.03$~K and $\sim0.15$~K respectively.

Case~(d) applies the hot-load $T_\mathrm{NS}(\nu)$ 4-NWP fit to the masked data.
The validator residual remains comparable to case~(c), but the antenna RMSE is larger
than in the scalar masked case. This is a stress-test limitation of the bad-calibrator subband. The
hot-load inversion divides the residual hot-load power by the relatively small
$X_\mathrm{NS}^\mathrm{hot}$ response, so small errors in the four fitted NWP terms or in
the masked-edge extrapolation are interpreted as structure in $T_\mathrm{NS}(\nu)$. When
that biased curve is applied to the antenna, where the noise-source coupling is much
larger, the error is amplified. Thus the hot-load route is useful for recovering a
frequency-dependent noise-source curve in the well-constrained full-band mock case, but it
is not the best-performing option for this deliberately poor calibrator set.

\section{Conclusions}
\label{sec:conclusions}

We have investigated the numerical stability of the REACH Bayesian noise wave calibration pipeline and identified a critical ill-conditioning problem that renders calibration solutions non-reproducible across different computing environments.
Our main findings and contributions are summarized as follows.

(i) Instability Diagnosis:
The calibration pipeline applied to the mock data produces condition numbers $\kappa(\mathbf{V}^*) \sim 10^{9}$--$10^{11}$ in the posterior covariance matrix, depending on the NumPy version used.
At this level of ill-conditioning, floating-point differences between software environments are amplified by many orders of magnitude, yielding qualitatively different calibration solutions from identical inputs.
This is a fundamental reproducibility failure that invalidates any scientific conclusions drawn from such solutions.

(ii) Four-noise-wave-parameter model and hot-load $T_\mathrm{NS}(\nu)$:
Fixing the excess noise temperature $T_\mathrm{NS}$ and fitting only the remaining four noise wave parameters $(T_\mathrm{unc}, T_\mathrm{cos}, T_\mathrm{sin}, T_\mathrm{L})$ eliminates the most degenerate direction in parameter space.
This reduces $\kappa(\mathbf{V}^*)$ from $\sim4.9\times10^{4}$ to $\sim60$ on the mock data and makes the solution reproducible across software environments.
A scalar fixed $T_\mathrm{NS}$, however, leaves a chromatic antenna residual when the true noise-source temperature varies across the band.
The iterative hot-load method developed here recovers a smooth $T_\mathrm{NS}(\nu)$ curve from the hot-load calibration spectrum and then keeps that curve fixed in the final 4-NWP fit.
On the mock dataset this reduces the binned antenna RMSE to $0.047$~K while preserving a validator RMSE of $0.015$~K, demonstrating that the data-driven hot-load route can recover the missing spectral information without reintroducing the original 5-NWP degeneracy.

(iii) Cable standing-wave collinearity masking:
A third source of instability arises when calibrators are connected through cables of non-negligible electrical length. Standing-waves produce localised spikes in $\kappa(\mathbf{X})$ at certain frequencies, reintroducing near-collinearity in the design matrix even after the 4-NWP degeneracy has been addressed.
Using a set of six calibrators with mixed 2-metre and 10-metre cable runs, we demonstrate this effect over $90$--$130$~MHz and show that masking a narrow window ($\pm1.5$~MHz) around each detected spike substantially reduces calibration residuals. The hot-load $T_\mathrm{NS}(\nu)$ update can be combined with the same masking strategy because the final calibration remains a fixed-$T_\mathrm{NS}$ 4-NWP fit, although in this deliberately poor calibrator set the recovered hot-load curve absorbs masked-edge modelling errors and worsens the antenna RMSE relative to the scalar masked fit.

Together, these results establish a physically motivated procedure for stabilising the REACH noise wave calibration: (1) use the 4-NWP model to remove the systematic $X_\mathrm{NS}/X_\mathrm{L}$ degeneracy; (2) replace the scalar noise-source approximation with an iterative hot-load estimate of $T_\mathrm{NS}(\nu)$ when the data support it; and (3) identify and mask frequency channels where cable standing-wave collinearities produce localised spikes in $\kappa(\mathbf{X})$.
Future work will extend this framework to automated spike detection in real data and to sub-band calibration, in which separate frequency sub-bands are calibrated independently \citep{monslave17}, each using the iterative hot-load $T_\mathrm{NS}(\nu)$ model.

More broadly, this work demonstrates that numerical stability should be treated as a first-class calibration requirement in global 21-cm experiments, alongside the more commonly discussed requirements of calibration accuracy and precision. The degeneracy between $\mathbf{X}_\mathrm{NS}$ and $\mathbf{X}_\mathrm{L}$ is inherent to the noise wave formalism, not specific to the REACH implementation, and similar ill-conditioning may affect any experiment that employs noise wave calibration with a comparable calibrator suite. We therefore recommend that future global 21-cm calibration pipelines include condition number monitoring as a standard diagnostic and consider physically motivated parameter reductions, such as the 4-NWP approach presented here, to ensure that calibration solutions remain reproducible and robust.

\section*{Acknowledgements}

SD acknowledges the Cambridge Trust and Isaac Newton Studentship for funding his PhD. SD also acknowledges
the useful discussions with the REACH team, especially with Prof. Anastasia Fialkov for her insightful comments 
on the project. 
AKD acknowledges was supported by the Cambridge Trust.
DA acknowledges the support of STFC.
HTJB acknowledges support from the Kavli Institute for Cosmology Cambridge and the Kavli Foundation.
CJK was supported by the Science and Technology Facilities Council grant number ST/V506606/1.
\section*{Data Availability}

The data underlying this article will be shared on reasonable request
to the corresponding author.


\bibliographystyle{mnras}
\bibliography{example} 


\bsp	
\label{lastpage}
\end{document}